\documentclass[11pt]{amsart}
\usepackage{geometry}                
\usepackage{graphicx}
\usepackage{amssymb}
\usepackage{epstopdf}
\usepackage[utf8]{inputenc}
\usepackage{hyperref}
\usepackage{tikz}
\usetikzlibrary{decorations.pathmorphing}
\usepackage{natbib}
\usepackage{algorithm,algpseudocode}
\usepackage{makecell}

\title{Deep Active Inference}
\author{Kai Ueltzhöffer}
\date{\today\\\\This is a pre-print of an article published in Biological Cybernetics. The final authenticated version is available online at: \url{https://doi.org/10.1007/s00422-018-0785-7}\\The author has archived a personal copy of the accepted manuscript at his personal homepage: \url{https://kaiu.me}}   

\begin{document}


\begin{abstract}
This work combines the free energy principle from cognitive neuroscience and the ensuing active inference dynamics with recent advances in variational inference on deep generative models and evolution strategies as efficient large-scale black-box optimisation technique, to introduce the "deep active inference" agent. This agent tries to minimize a variational free energy bound on the average surprise of its sensations, which is motivated by a homeostatic argument. It does so by changing the parameters of its generative model, together with a variational density approximating the posterior distribution over latent variables, given its observations, and by acting on its environment to actively sample input that is likely under its generative model. The internal dynamics of the agent are implemented using deep neural networks, as used in machine learning, and recurrent dynamics, making the deep active inference agent a scalable and very flexible class of active inference agents. Using the mountaincar problem, we show how goal-directed behaviour can be implemented by defining sensible prior expectations on the latent states in the agent's model, that it will try to fulfil. Furthermore, we show that the deep active inference agent can learn a generative model of the environment, which can be sampled from to understand the agent's beliefs about the environment and its interaction with it.
\end{abstract}

\maketitle

\section{Introduction}

Active Inference \citep{FristonKilnerHarrison2006, FristonAl2010, Friston2012} is a normative theory of brain function derived from the properties required by active agents to survive in dynamic, fluctuating environments. This theory is able to account for many aspects of action and perception as well as anatomic and physiologic features of the brain on one hand \citep{Brown2012, Friston2005,Friston2011,Friston2009,Schwartenbeck2015, Adams2013}, and encompasses many formal theories about brain function on the other \citep{Friston2010}. 
In terms of its functional form it rests on the minimisation of an upper variational bound on the agents average surprise. In this way it is formally very similar to state of the art algorithms for variational inference in deep generative models \citep{RezendeMohamedWierstra2014, KingmaWelling2014, Chung2015}. However, optimising this bound for active agents introduces a dependency on the true dynamics of the world, to which the agent usually does not have access, and whose true functional form does not have to coincide with the functional form of the agent's generative model. Here we solve these problems using deep neural networks \citep{Lecun2015} and recurrent neural networks \citep{Karpathy2015} as flexible function approximators, which allow the agent's generative model to learn a good approximation of the true world dynamics. Futhermore we apply evolution strategies \citep{Salimans2017} to estimate gradients on the variational bound, averaged over a population of agents. This formalism allows to obtain gradient estimates even for non-differentiable objective functions, which is the case here; since the agent does neither know the equations of motions of the world, nor its partial derivatives. 

In this way, our approach pairs active inference with state of the art machine learning techniques to create agents that can successfully reach goals in complex environments while simultaneously building a generative model of themselves and their surroundings. In this work we want to lay out the basic form of these so called "Deep Active Inference" agents and illustrate their dynamics using a simple, well known problem from reinforcement learning, namely the mountain car problem \citep{Moore1991}. By utilising models and optimisation techniques that have been applied successfully to real-world data and large-scale problems \citep{RezendeMohamedWierstra2014, KingmaWelling2014, Chung2015, Salimans2017}, our agent can be scaled to complex and rich environments. 
With this paper we publish the full implementation of the resulting Deep Active Inference agent, together with all scripts used to create the figures in this publication at \url{https://www.github.com/kaiu85/deepAI_paper}.

In section 2 we will briefly recapitulate the active inference principle. In section 3 we will describe the mountain car environment and introduce a deep active inference agent that is able to solve this problem. In section 4 we will show that the agent can solve the mountain car problem while simultaneously learning a generative model of its environment. In section 5 we will discuss possible further directions of this approach and its relation to other approaches.

\section{Active Inference}

\label{sec_free_energy}

Active Inference \citep{FristonKilnerHarrison2006, FristonAl2010, Friston2012} rests on the basic assumption that any agent in exchange with a dynamic, fluctuating environment has to keep certain inner parameters within a well defined range. Otherwise, it would sooner or later encounter a phase transition due to which it would loose its defining characteristics and therefore disappear. Thus, the agent must restrict itself to a small volume in its state space. This can be formalised using the entropy of the probability distribution $p(s^*)$ of finding the agent in a given state $s^*$ of its state space $ S^*$:
$$ H(S^*) =  \int_{s^*\in S^*} \left( -\ln p(s^*)\right) p(s^*) \,\mathrm{d}s^*$$
By minimising this entropy, an agent can counter dispersive effects from its environment and maintain a stable identity in a fluctuating environment.

However, an agent does not have direct access to an objective measurement of its current state. Instead it only perceives itself and the world around it via its sensory epthelia. This can be described by a potentially noisy sensory mapping $o = g(s)$ from states $s$ to sensations $o$. Defining the sensory entropy
$$ H(O) = \int_{o\in O} \left( -\ln p(o)\right) p(o) \,\mathrm{d}o $$
over the space $O$ of all possible observations of an agent, one can derive the following inequality in the absence of sensory noise \citep{FristonAl2010}
$$H(O) \geq H(S^*) + \int_{s^*\in S^*} p(s^*) \ln \left| g_{s^*} \right| \mathrm{d}s$$

Agents, whose sensory organs did not have a good mapping of relevant physical states to appropriate sensory inputs, would not last very long.
So the mapping between the agents true states and its sensations is assumed to have an almost constant sensitivity, in terms of the determinant of the Jacobian $\left| g_{s^*} \right|$ over the encountered range of states $s$. This makes the last term approximately constant and allows upper bounding the entropy of the agent's distribution on state space by the entropy of its sensory states plus this constant term \citep{FristonAl2010}. Thus, to ensure keep its physiological variables within well-defined bounds, an agent has to minimize its sensory entropy $H(O)$ \footnote{In more general formulations of active inference, the assumption that the mapping between hidden states and outcomes is constant can be relaxed \citep{FristonEpistemicValue}.}

Assuming ergodicity, i.e. the equivalence of time- and ensemble-averages, one can write the sensory entropy as
$$ H(O) = \lim\limits_{T \rightarrow \infty} - \frac{1}{T} \int_0^T \ln p(o(t)) \,\mathrm{d}t$$

From the calculus of variations it follows that an agent can minimize its sensory entropy by minimising its sensory surprise $ - \ln p(o(t))$ at all times, in terms of the following Euler-Lagrange-equation:
$$ \nabla_o \left( - \ln p(o(t)) \right) = 0$$

To be able to efficiently do this, our agent needs a statistical model of its sensory inputs, to evaluate $p(o)$. Since the world in which we live is hierarchically organised, dynamic, and features a lot of complex noise, we assume that the agent's model is a deep, recurrent, latent variable model \citep{Ashby1970}. Furthermore we assume that this model is generative, using the observation that we are able to imagine certain situations and perceptions (like the image of a blue sky over a desert landscape) without actually experiencing or having experienced them. Thus, we work with a generative model $ p_\theta(o,s)$ of sensory observations $o$ and latent variables $s$, that represent the hidden true states of the world, which we can factorise into a likelihood function $ p_\theta(o|s)$ and a prior on the states $ p_\theta(s)$:
$$ p_\theta(o,s) = p_\theta(o|s) p_\theta(s)$$

The set $ \theta$ comprises the slowly changing parameters that the agent can change to improve its model of the world. In the brain this might be the pattern and strength of synapses. Given this factorisation, to minimize surprise, the agent has to solve the hard task of calculating
$$ p_\theta(o) = \int p_\theta(o|s)p_\theta(s) \,\mathrm{d}s$$

by marginalising over all possible states that could lead to a given observation. As the dimensionality of the latent state space $ S$ can be very high, this integral is extremely hard to solve. Therefore a further assumption of the free energy principle is, that agents do not try to minimize the surprise $ -\ln p_\theta(o)$ directly, but rather minimize an upper bound, which is a lot simpler to calculate.

Using the fact that the Kullback-Leibler (KL) Divergence

$$ D_{\mathrm{KL}}(p_a(x)||p_b(x)) = \int_{x \in X} \ln\left( \frac{p_a(x)}{p_b(x)}\right)p_a(x) \,\mathrm{d}x$$

between two arbitrary distributions $p_a(x)$ and $ p_b(x)$ with a shared support on a common space $ X$ is always greater or equal to zero, and equal to zero if and only if the two distributions are equal, we can define the variational free energy as:

%
%

$$ F(o,\theta, u) = -\ln p_\theta(o) + D_{\mathrm{KL}}(q_u(s)||p(s|o)) \geq   -\ln p_\theta(o)$$

Here $ q_u(s)$ is an arbitrary, so called variational density over the space of hidden states $ s$, which belongs to a family of distributions parameterised by a time-dependent, i.e. fast changing, parameter set $ u$. If $ q_u(s)$ was a diagonal Gaussian, $ u = \left\{\mu, \sigma\right\}$ would be the corresponding means and standard-deviations. This parameter set can be encoded by the internal states of our agent, e.g. by the neural activity (i.e. the firing pattern of neurons) in its brain. Thus, the upper bound $ F(o,\theta, u)$ now only depends on quantities to which our agent has direct access. Namely the states of its sensory organs $ o$, the synapses encoding its generative model of the world $ \theta$ and the neural activity representing the sufficient statistics $ u$ of the variational density.

Using the definition of the Kullback-Leibler divergence, the linearity of the integral, Bayes' rule, and manipulation of logarithms, one can derive the following equivalent forms of the free energy functional:

$$
\begin{array}{rcl}
F(o,\theta, u) & = & -\ln p_\theta(o) + D_{\mathrm{KL}}(q_u(s)||p_\theta(s|o)) \\
& = & \left< -\ln p_\theta(o,s)  \right>_{q_u(s)} - \left< - q_u(s)  \right>_{q_u(s)}\\
& = & \left< -\ln p_\theta(o|s)  \right>_{q_u(s)} + D_{\mathrm{KL}}(q_u(s)||p_\theta(s)) \\
\end{array}
$$

Here $ \left< f(s) \right>_{q_u(s)}$ means calculating the expectation value of $ f(s)$ with respect to the variational density $ q_u(s)$.

If the agent was unable to change its environment,  i.e. just a passive observer of the world around it, the only thing it could do to minimize $ F$ would be to change the parameters $ \theta$ of its generative model and the sufficient statistics $ u$ of its inner representation. Looking at the first form of the free energy, optimizing $ u$ would correspond to minimizing the Kullback-Leibler divergence between the variational distribution $ q_u (s)$ and the true posterior distribution $ p(s|o)$, i.e. the probability over hidden states $ s$ given the observations $ o$. Thus, the variational density can be seen as an approximation of the true posterior density. Therefore, by minimising $F$, the agent automatically acquires a probabilistic representation of an approximate posterior on the hidden states of the world, given its sensory input. The optimization of the sufficient statistics $ u$ of the variational density $ q$ is therefore what we call "perception". As $ q$ has to be optimized on a fast timescale, quickly changing with the sensory input $ o$, it is likely represented in terms of neural activity. This might explain hallmarks of Bayesian computations and probabilistic representations of sensory stimuli in the brain \citep{Ernst2002, Alais2004, Knill2004, Moreno-Bote2011, Berkes2011, Haefner2016}. 

As the variational free energy upper bounds the surprise $ -\ln p_\theta(o)$ , minimising free energy with respect to the parameters $ \theta$ of the generative model will simultaneously maximise the evidence $ p_\theta(o)$ for the agent's generative model of the world. The resulting optimisation of the parameters $ \theta$ of the generative model is what we call perceptual learning and what might be implemented by changing the synaptic connections between neurons in the brain. The second form is given only to demonstrate, where the name "free energy" originated from, since its form is very similar to the Helmholtz free energy of the canonical ensemble in statistical physics.

The core idea of active inference is now to equip the agent with actuators, that allow it to actively change the state of its environment. Thereby the sensory observations $ o$ become a function of the current and past states $ a$ of the agents actuators ("muscles"), via their influence on the state of the world generating the sensory inputs. Now the agent can not only minimize the free energy bound by learning (i.e. optimising the parameters of its generative model) and perception (i.e. optimising the sufficient statistics of the variational density), but also by changing the observations it makes.

Considering the third form of the variational free energy

$$ F(o,\theta, u) = \left< -\ln p_\theta(o|s)  \right>_{q_u(s)} + D_{\mathrm{KL}}(q_u(s)||p_\theta(s)) $$

we see that our agent can minimize it by seeking out observations $ o$ that have a high likelihood $ p_\theta(o|s)$  under its own generative model of the world, averaged over its approximate posterior of the state of the world. Thus, the agent will seek out states that conform to its expectations, i.e. that have a high likelihood under its generative model of the world. So one can encode specific goals in the priors of the generative model of the agent: assigning a higher a priori probability to a specific state, the agent will try to seek out this state more often. The first term, which our agent will maximise using its actions, is also called accuracy in Bayesian statistics, while the second term is known as complexity. Complexity measures the degree to which posteriors have to be adjusted in relation to priors to provide an accurate account of sensory data or outcomes.

This triple optimisation problem can be formalised using the following dynamics for the parameters $ \theta$ of an agent's generative model, its internal states $ u$, encoding the sufficient statistics of the variational density $ q$, and the states of its actuators $ a$:

$$
\left(\theta, u, a\right) = \underset{(\tilde{\theta}, \tilde{u}, \tilde{a})}{\mathrm{arg\,min}}F(o(\tilde{a}),\tilde{\theta},\tilde{u})
$$

In this paper, we will consider the variational density or beliefs over hidden states of the world. In more general formulations, this density would cover both the hidden states and the parameters of the generative model. Usually, these beliefs are factorised so that there is a mean field approximation to the true posterior estimates of (time varying) states and (time invariant) parameters \citep{Friston2008}.

\section{Methods}

In this section, we introduce an agent which uses recent advances in stochastic optimisation of deep generative models \citep{KingmaWelling2014, RezendeMohamedWierstra2014, Chung2015} and evolution strategies as efficient, scalable optimisation algorithm for non-differentiable objective functions \citep{Salimans2017} to implement active inference. We show its ability to reach a specific goal while concurrently building a generative model of its environment, using the well-known mountain car problem \citep{Moore1991}.
But we first start with some basics that are well known in the deep learning community, but which - due to similar but differently used nomenclature - might lead to confusion among computational neuroscientists.

\subsection{Deep Neural Networks}

When we talk about deep neural networks, we use this term in the sense of the deep learning literature \citep{Lecun2015}. In this sense, neural networks are nothing but very flexible function approximators. An excellent and comprehensive introduction to the state of the art in deep learning is given in \cite{Goodfellow2016}.

\subsubsection{Fully Connected Layer} Deep neural networks are composed of individual layers, which are functions $f_\theta: \mathbb{R}^{d_1} \rightarrow \mathbb{R}^{d_2}, x\mapsto f_\theta(x)$. Here $d_1$ represents the number of input neurons, i.e. the dimensionality of the input space, and $d_2$ represents the number of output neurons, i.e. the dimensionality of the output space. The subscript $\theta$ represents the parameters of the function, which can be tuned to approximate or optimise a given objective function. 

The canonical functional form of a so called \textit{fully connected} layer is:

$$f_\theta(x) = \textrm{h}\left( \theta_w x + \theta_b \right)$$

Here $\textrm{h}: \mathbb{R} \rightarrow \mathbb{R}$ is an \textit{elementwise} nonlinear transfer function. This could for example be a $\mathrm{tanh}$, $\mathrm{sigmoid}$ or a so called rectified linear function $\mathrm{relu(x)} = \max(0,x)$. The set of parameters $\theta = \{\theta_w, \theta_b\}$ consists of the $d_2 \times d_1$ matrix $\theta_w$, called the weight matrix or just weights, and the bias vector $\theta_b \in \mathbb{R}^{d_2}$. This functional form is loosely inspired by the response function of ensembles of simple point neurons (e.g. the mean-field equation for the average firing rate of a large population of leaky-integrate-and-fire neurons as function of the mean inputs to the population, c.f. equations 12 - 15 in \cite{WongWang2006} ). These response functions convert the weighted sum of inputs currents (approximated by the weighted sum of the firing rates of the presynaptic neurons) to the firing rate of the modelled population by a nonlinear, thresholded activation function, which is represented here using the bias parameters $\theta_b$ and the nonlinear transfer function $\mathrm{h}$.

\subsubsection{Deep Feedforward Networks} A deep feedforward network just consists of a stack of individual layers $f_1, f_2, f_3, ..., f_{n-1}, f_n$, that are applied to the input $x$ sequentially to generate an output $y$:

$$ y = f_n(f_{n-1}( ... f_3(f_2(f_1(x))) ... )) = f_n \circ f_{n-1} \circ ... \circ f_3 \circ f_2 \circ f_1 (x) $$

The output of the function $f_n$ is called the output layer, the input $x$ is called the input layer, the intermediate values $f_1(x), f_2(f_1(x)), ..., f_{n-1}(...x)$ are called the hidden layers, since they are not explicitly constrained by the target or objective function. The finite dimensions of the intermediate outputs $f_1(x), f_2(f_1(x)), ..., f_{n-1}(...x)$ can be regarded as the number of hidden neurons in each layer.

In contrast to previous approaches to active inference \citep{Friston2010} deep feedforward networks allow us to specify agents, whose generative models of their environment do not need to be from the same model family as the true generative process. This is due to the flexibility of deep feedforward networks. It was shown that a feedforward network with a linear output layer and at least one hidden layer with a nonlinear activation function, such as a $\mathrm{tanh}$ or $\mathrm{sigmoid}$ activation function can approximate any Borel measurable function from one finite-dimensional space to another with arbitrary accuracy, given a sufficient dimensionality (i.e. a sufficient number of neurons) of the hidden layer \citep{Hornik1989}.

\subsubsection{Recurrent Neural Networks}

To enable a learning system to build a memory of its previous actions and experiences, a simple and elegant way is to enhance a neural network

$$ y_t = f_\theta(x_t) $$

by feeding its output at the previous timestep back as an additional input, i.e. 

$$ y_t = f_\theta(x_t, y_{t-1}) $$

Although there is a wide variety of more complex and powerful architectures (c.f. \cite{Karpathy2015, Goodfellow2016}), this simple, prototypical recurrent neural network was recently shown to be able to learn long-range temporal dependencies, given a sensible initialization \citep{Hinton2015}. We are not using this deterministic type of recurrent neural network directly, but we use the basic idea by feeding back samples from a distribution on the current network state as inputs to calculate the next network state. Similar to the approximation theorem for deep feedforward networks, it was shown that recurrent neural networks are able to learn arbitrary algorithms, as they are turing complete \citep{Siegelmann1995}.

\subsection{The Environment}

\begin{figure}
\includegraphics[width=0.5\paperwidth]{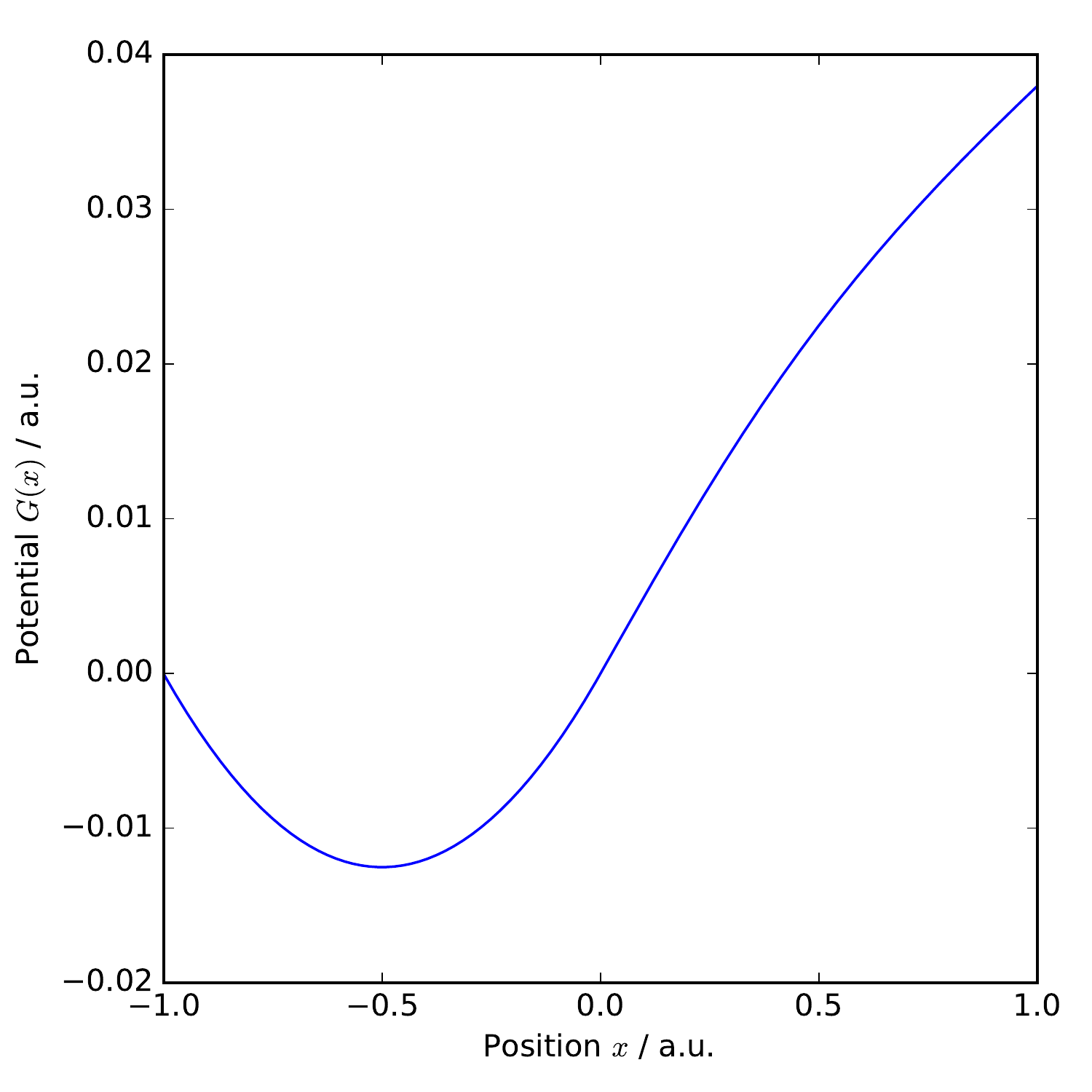}
\caption{Potential of the mountain car problem. The agent is a small cart starting at the bottom of the valley at $x=-0.5$ }
\label{mountaincar_potential}
\end{figure}

The agent will act in a discrete time version of the mountain car world \citep{Moore1991}. It will start at the bottom $x = -0.5$ of a potential landscape $G(x)$, which is shown in figure \ref{mountaincar_potential}.

The agent itself is a small car on this landscape, whose true, physical state $\mathbf{s}_t^* = \left(a_t, x_t, v_t \right) $ is given by its current position $ x_t$, its velocity $ v_t$,  and the current state $a_t$ of its effector organs. In the case of humans this would be the state of our muscles, here the state $a_t$ describes the car's steering direction and the throttle of the car's engine. 
The dynamics of the environment are given by a set of update equations, which can be summarised symbolically by
$$(x_{t+1},v_{t+1}) = \mathbf{R}(x_t, v_t, a_{t+1}) $$
Note that we use bold face to denote vectors. 

\begin{figure}
\includegraphics[width=0.5\paperwidth]{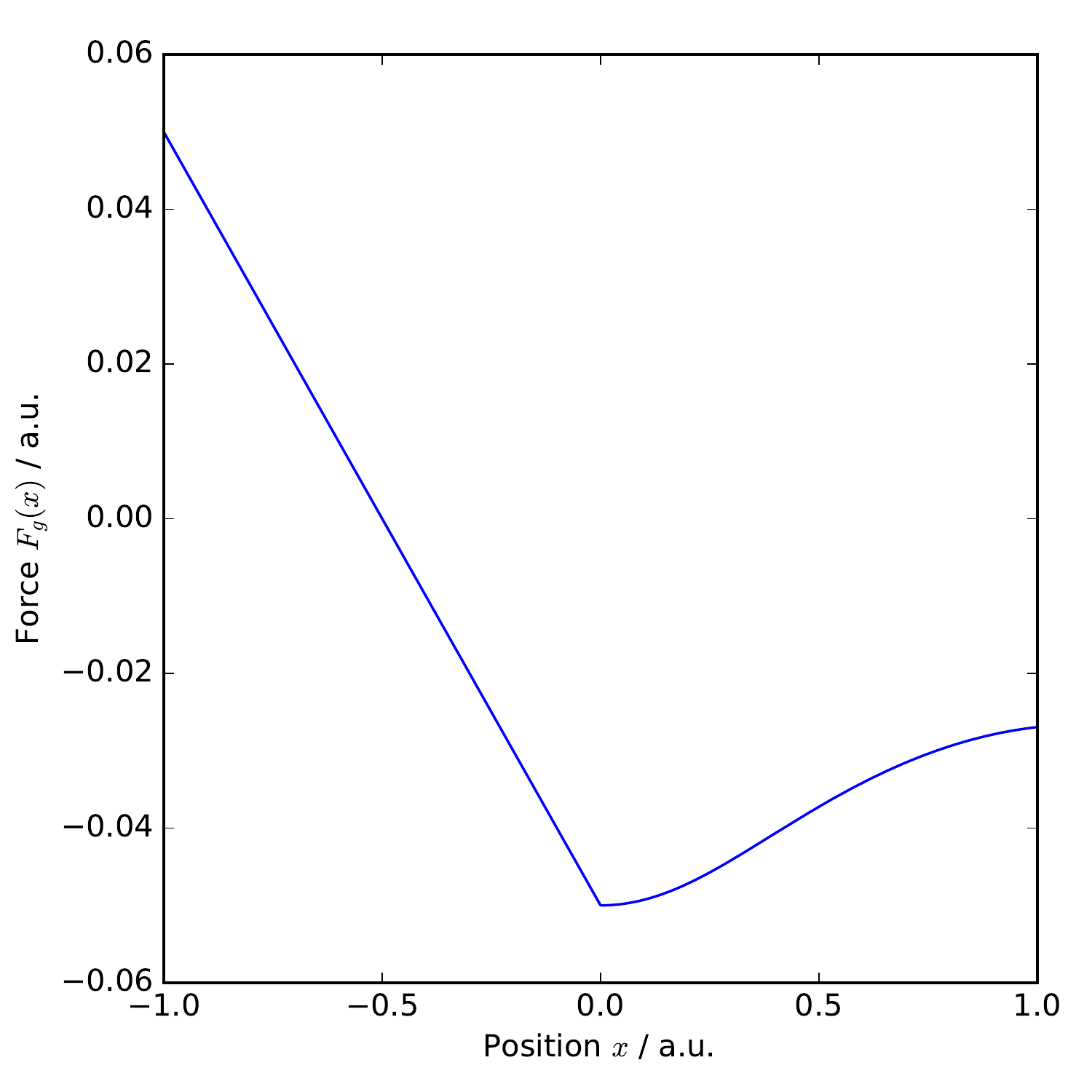}
\caption{ Downhill force $F_g$ due to the slope of the potential landscape.}
\label{mountaincar_force}
\end{figure}

This equation can be decomposed into a set of equations, as multiple forces will act on the agent. The downhill force $F_g$ due to the shape of the landscape depends on its position $x$. It is given by
$$F_g(x)  = -\frac{\partial}{\partial x}G(x) = \begin{cases} 0.05\left( -2x - 1 \right), & x < 0 \\ 0.05 \left( -(1+5x^2)^{-1/2}-x^2(1 + 5x^2)^{-3/2}-x^4/16 \right), & x \geq 0 \end{cases} $$ and shown in figure \ref{mountaincar_force}.

The agent's motor can generate a force $F_a$, depending on its current action state $a$ 
$$F_a(a) = 0.03 \tanh (a) $$
As mentioned above, the action state controls the steering direction (positive or negative) and the throttle of the engine. However, the absolute force that the engine can generate is limited to the interval $(-0.03, 0.03)$ due to the $\tanh$ function. 

The laminar friction force $F_f$ depends linearly on the agent's current velocity $v$ 
$$F_f(v) = - 0.25v $$

Thus, the total force action on the agent is
$$F_\mathrm{total}(x_t, a_{t+1}, v_t) = F_g(x_t) + F_a(a_{t+1}) + F_f(v_t) $$

This leads to the following update equations for velocity
$$v_{t+1} = v_t + F_\mathrm{total}(x_t, a_{t+1}, v_t) $$
and position
$$x_{t+1} = x_t + v_{t+1} $$

Initially, the agent is resting at the bottom of the landscape, i.e. $\mathbf{s}^*_0 = \left( x_0, v_0 \right) = \left( -0.5, 0.0 \right)$.

We will later set the agent the goal of reaching and staying at $x = 1.0$. However, due to the shape of the potential landscape and the resulting force $F_g(x)$, as shown in figure \ref{mountaincar_force}, we notice that the landscape gets very steep at $x = 0.0$.
As the force generated by the motor is limited to the interval $(-0.03, 0.03)$, the agent is not strong enough to climb the slope at $x=0$, which results in a downhill force of $F_g(0) = -0.05$, without some additional momentum. Thus, to overcome this slope the agent has to move uphill to the left, \textit{away} from its target position, at first. In this way, it can acquire the required additional momentum, which allows it to overcome the opposite slope. In this way, the mountaincar environment - although very simple - is not completely trivial, as the agent has to learn that a direct approach of the target position will not succeed and it needs to acquire a strategy that initially leads away from its target.

\subsection{The Deep Active Inference Agent}

We will now describe an agent that follows the active inference principle laid out in section \ref{sec_free_energy} \citep{FristonKilnerHarrison2006, FristonAl2010}. It is adapted to the mountain car environment by its sensory inputs and motor outputs and its inner workings are implemented and optimised using recent advances in deep learning \citep{KingmaWelling2014, RezendeMohamedWierstra2014, Chung2015, Salimans2017}.

\subsubsection{Sensory Inputs}

Our agent has a noisy sense $ o_{x,t}$ of its real position $ x_t$

$$ p(o_{x,t} | x_t) = \mathrm{N}(o_{x,t} ; x_t, 0.01)$$

Here $\mathrm{N}(x ; \mu, \sigma)$ denotes a Gaussian probability density over $x$ with mean $\mu$ and standard-deviation $\sigma$. To show that the agent can indeed learn a complex, nonlinear, generative model of its sensory input, we add another sensory channel with a nonlinear, non-bijective transformation of $ x_t$:

$$ o_{h,t} = \exp\left(-\frac{(x_t-1.0)^2}{2\cdot 0.3^2}\right) $$


Note that the agent has no direct measurement of its current velocity $ v_t$, i.e. it has to infer it from the sequence of its sensory inputs, which all just depend on $ x_t$.

To see if the agent understands its own actions, we also equip it with a proprioceptive sensory channel, allowing it to observe its action state $ a_{t}$:

$$ o_{a,t} = a_{t} $$


Note that having a direct sense of its action state $a_t$ is not necessary for an active inference agent to successfully control its environment (c.f. supplementary figures 9-11). E.g. the reflex arcs that we use to increase the likelihood of our sensory inputs, given our generative model of the world, feature their own, closed loop neural dynamics at the level of the spinal cord. So we do not (and do not have to) have direct access to the action states of our muscles, when we just lift our arm by expecting it to lift. However, adding this channel will allow us later to directly sample the agents proprioceptive sensations from its generative model, to check if it understands its own action on the environment.

\subsubsection{The Agent's Generative Model of the Environment}

First, the agent has a prior over the hidden states of the world $ \mathbf{s}$:

$$ p_\theta(\mathbf{s}_1, \mathbf{s}_2, ..., \mathbf{s}_T) = \prod_{t=1}^T p_\theta (\mathbf{s}_t|\mathbf{s}_{t+1})$$

with initial state $ \mathbf{s}_0 = (0,...,0)$. Our agent will use $d_s = 10$ neurons to represent the current state of the world.

This factorisation means that the next state $ \mathbf{s}_{t+1}$ in the agents generative model only depends on the current state $ \mathbf{s}_t$ of its inner model of the world. Thus, the transition distribution $ p_\theta (\mathbf{s}_t|\mathbf{s}_{t+1})$ encodes the agent's model of the dynamics of the world, i.e. what it has learned about the laws of motion of its environment and its influence on it.

In our concrete implementation, we model the distribution of $ \mathbf{s}_{t+1}$ as diagonal Gaussian, where the means and standard deviations are calculated from the current state $ \mathbf{s}_t$ using neural networks. I.e.

$$ p_\theta(\mathbf{s}_{t+1}|\mathbf{s}_t) = \mathrm{N}(\mathbf{s}_{t+1}; \boldsymbol{\mu}^t_\theta(s_t), \boldsymbol{\sigma}^t_\theta(s_t))$$

We use a fully connected single layer network with a $\tanh$ nonlinearity to calculate the means $ \boldsymbol{\mu}^t_\theta(s_t)$ and another fully connected single layer network with $\mathrm{softplus(x)} = \ln(1 + e^x)$ as nonlinear transfer function to calculate the standard deviations $ \boldsymbol{\sigma}^t_\theta(s_t)$.
We use $ \theta$ to encompass all parameters of the generative model, that we are going to optimise. In practice, this means all the weights and bias parameters of the neural networks to calculate the means and standard-deviations. The use of diagonal Gaussians in the prior leads to hidden, abstract representations within which independent causes of observations are well separated.

Similarly, the likelihood functions for each of the three observables also factorise
\begin{multline*}
p_\theta(o_{x,1}, ..., o_{x,T},o_{h,1}, ..., o_{h,T},o_{a,1}, ..., o_{a,t}|\mathbf{s}_1, ...., \mathbf{s}_T) \\
= \prod_{t=0}^T p_\theta (o_{x,t} | \mathbf{s}_t) \prod_{t=0}^T p_\theta (o_{h,t} | \mathbf{s}_t) \prod_{t=0}^T p_\theta (o_{a,t} | \mathbf{s}_t)
\end{multline*}
So the likelihood of each observable $ o_{x,t}, o_{a,t}, o_{h,t}$ for a given time $ t$ only depends on the current state $ s_t$. We again use Gaussian distributions, to obtain
\begin{align*} 
p_\theta(o_{x,t}|\mathbf{s}_t) & = & \mathrm{N}(o_t; \boldsymbol{\mu}^{x}_\theta(\mathbf{s}_t), \boldsymbol{\sigma}^{x}_\theta(\mathbf{s}_t)) \\
p_\theta(o_{a,t}|\mathbf{s}_t) & = & \mathrm{N}(o_t; \boldsymbol{\mu}^{a}_\theta(\mathbf{s}_t), \boldsymbol{\sigma}^{a}_\theta(\mathbf{s}_t)) \\
p_\theta(o_{h,t}|\mathbf{s}_t) & = & \mathrm{N}(o_t; \boldsymbol{\mu}^{h}_\theta(\mathbf{s}_t), \boldsymbol{\sigma}^{h}_\theta(\mathbf{s}_t))
\end{align*}
We calculate the sufficient statistics of these Gaussian distributions from the state $\mathbf{s}_t$ using a deep feedforward network with three hidden layers of $d_h = 10$ neurons each, using $\tanh$ as nonlinear transfer function. We use a linear output layer to calculate the means of the Gaussian variables and a second output layer with $\mathrm{softplus(x)} = \ln(1 + e^x)$ as nonlinear transfer function to calculate the standard deviations.  Although this structure is on the first glance different from the hierarchical dynamical models developed by \cite{Friston2008}, the deep and nonlinear structure of the feedforward network also allows for structured noise to enter at different levels of this hierarchy.

Once the agent has acquired a generative model of its sensory inputs, by optimising the parameters $ \theta$, one can sample from this model by propagating the prior on the state space and then using the learned likelihood function to generate samples, as described in algorithm \ref{sampling_from_gm}. Sampling many processes ($n_p \approx 10^2-10^4$) in parallel yields a good approximation, although only a single sample is drawn from the prior density and from each factor of the likelihood function per individual process and per timestep.

\begin{algorithm}[t]
\caption{Sampling from the agents generative model. Concrete samples and values of a variable marked using a hat, e.g. $\hat{\mathbf{s}}_t$.}
\algnewcommand\algorithmicforeach{\textbf{for each}}
\algdef{S}[FOR]{ForEach}[1]{\algorithmicforeach\ #1\ \algorithmicdo}
\begin{algorithmic}
\State Initialize $n_p$ processes with $\hat{\mathbf{s}}_0 = (0,...,0)$.
\ForEach {Process}
\For{$t=1,...,T$}
\State Draw a single sample $ \hat{\mathbf{s}}_t$ from $ p_\theta(\mathbf{s}_t | \hat{\mathbf{s}}_{t-1})$
\State Sample single observations $ \hat{o}_{x/a/h}$ from each likelihood $ p_\theta(o_{x/a/h,t} | \hat{\mathbf{s}}_t)$
\State Carry $ \hat{s}_t$ over to the next timestep.
\EndFor
\EndFor
\end{algorithmic}
\label{sampling_from_gm}
\end{algorithm}

\subsubsection{Variational Density}

Following \cite{KingmaWelling2014, RezendeMohamedWierstra2014} we do not explicitly represent the sufficient statistics of the variational posterior at every time step. This would require an additional, costly optimisation of these variational parameters at each individual time step. Instead we use an inference network, which approximates the dependency of the variational posterior on the previous state and the current sensory input and which is parameterised by time-invariant parameters. This allows us to learn these parameters together with the parameters of the generative model and the action function (c.f. section \ref{sec_action_states}), and also allows for very fast inference later on. We use the following factorisation for this approximation of the variational density on the states $\mathbf{s}$, given the agents observations o:
\begin{multline*}
q_\theta(\mathbf{s}_1, ..., \mathbf{s}_T | o_{x,1}, ..., o_{x,T},o_{h,1}, ..., o_{h,T}, o_{a,1}, ..., o_{a,T}) \\
= \prod_{t=1}^T q_\theta(\mathbf{s}_t | \mathbf{s}_{t-1}, o_{x,t}, o_{h,t}, o_{a,t}) 
\end{multline*}
with initial state $\mathbf{s}_0 = (0,...,0)$, before the agent has interacted with the environment. We again use diagonal Gaussians
$$
q_\theta(\mathbf{s}_t | \mathbf{s}_{t-1}, o_{x,t}, o_{h,t}, o_{a,t}) \\
 = \mathrm{N}(\mathbf{s}_t; \boldsymbol{\mu}^q_\theta(\mathbf{s}_{t-1}, o_{x,t}, o_{h,t}, o_{a,t}), \boldsymbol{\sigma}^q_\theta(\mathbf{s}_{t-1}, o_{x,t}, o_{h,t}, o_{a,t}))
$$
The sufficient statistics are calculated using a deep feedforward network with two hidden layers of $d_h = 10$ neurons each, using $\tanh$ nonlinearities. The means are again calculated using a linear, and the standard deviations using a $\mathrm{softplus}$ output layer.
While the use of diagonal Gaussians in the prior leads to hidden, abstract representations within which independent causes of observations are optimally separated, i.e. which has favourable properties, here this choice is just due to practical considerations. The fact that we choose both the variational density and the prior density as diagonal Gaussians, will later allow us to use a closed formula to calculate the Kullback-Leibler-Divergence between the densities. However, if a more flexible posterior is required, normalizing flows allow a series of nonlinear transformations of the diagonal Gaussian used for the variational density, by which it can approximate very complex and multimodal posterior distributions \citep{RezendeMohamed2015, KingmaIAF, TomczakHHFlow}.

This device – of learning the mapping between inputs and the sufficient statistics of (approximate) posterior over hidden states – is known as amortisation. It has the great advantage of being extremely efficient and fast. On the other hand, it assumes that the mapping can be approximated with a static non-linear function that can be learned with a neural network. This somewhat limits the context sensitivity of active inference schemes, depending upon the parameterisation of the mapping. In short, amortisation enables one to convert a deep inference or deep deconvolution problem into a deep learning problem – by finding a static non-linear mapping between (time varying) inputs and (approximate) posterior beliefs about the states generating those inputs.

\subsubsection{Action States}
\label{sec_action_states}

In general, the action state $ a_t$ of the agent could be minimised directly at each timestep, thereby minimizing the free energy by driving the sensory input (which depends on the action state via the dynamics of the world) towards the agents expectations, in terms of its likelihood function. However, this would require a costly optimisation for each timestep (and every single simulated process). Thus, we use the same rationale as \cite{KingmaWelling2014, RezendeMohamedWierstra2014} used for the variational density $ q$, and approximate the complex dependence of the action state on the agent's current estimate of the world (and via this on the true state of the world) by fixed, but very flexible functions, i.e. deep neural networks. This yields an explicit functional dependency
$$ p_\theta(a_{t+1} | \mathbf{s}_t)$$
whose parameters we include, together with the parameters of the generative model and the variational density, to the set of parameters $\theta$ that we will optimise.

We use a one-dimensional Gaussian form
$$ p_\theta(a_{t+1} | \mathbf{s}_t) = \mathrm{N}(a_{t+1}; \mu_\theta^a(\mathbf{s}_t), \sigma_\theta^a(\mathbf{s}_t) ) $$
whose sufficient statistics are calculated using a deep feedforward network with one fully connected hidden layers of $d_h = 10$ neurons, a linear output layer for the mean, and a $\mathrm{softplus}$ output layer for the standard deviation.

We now can just optimise the time-invariant parameters $ \theta$ of these neural networks together with the parameters of the generative model and the variational density. This approximation makes learning and propagating the agent very fast, allowing for efficient simultaneous learning of the generative model, the variational density and the action function.

The approximation of both, the sufficient statistics of the variational density $ q$ and the action states $ a$ by deep neural networks is the reason why we call this class of agents \textit{Deep} Active Inference agents.

\begin{figure}
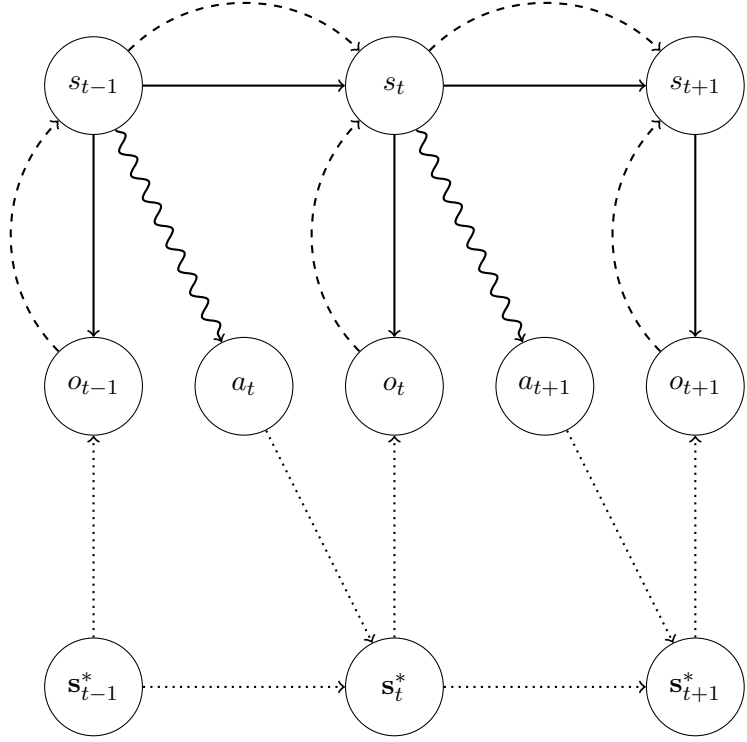


\tikzstyle{n}=[circle, draw=black, minimum size=1.3cm]
\tikzstyle{g}=[->,thick]
\tikzstyle{q}=[->,thick,dashed,bend left=45]
\tikzstyle{w}=[->,thick,dotted]
\tikzstyle{a}=[->,thick,decorate,decoration=snake]

\tikz {
	\node (stm1) at (0,4) [n]  {$s_{t-1}$};
	\node (st) at (4,4) [n] {$s_t$};
	\node (stp1) at (8,4) [n] {$s_{t+1}$};
	\node (otm1) at (0,0) [n]  {$o_{t-1}$};
	\node (ot) at (4,0) [n] {$o_t$};
	\node (otp1) at (8,0) [n] {$o_{t+1}$};
	\node (atm1) at (2,0) [n]  {$a_{t}$};
	\node (at) at (6,0) [n] {$a_{t+1}$};
	\node (xtm1) at (0,-4) [n]  {$\mathbf{s}^*_{t-1}$};
	\node (xt) at (4,-4) [n] {$\mathbf{s}^*_t$};
	\node (xtp1) at (8,-4) [n] {$\mathbf{s}^*_{t+1}$};
	\draw 
		  (stm1) edge [g] (st) 
	          (st) edge [g] (stp1)
	          (stm1) edge [g] (otm1)
	          (st) edge [g] (ot)
	          (stp1) edge [g] (otp1)
	          
	          (stm1) edge [q] (st)
	          (st) edge [q] (stp1) 
	          (ot) edge [q] (st)
	          (otm1) edge [q] (stm1)
	          (otp1) edge [q] (stp1)
	          
	          (xtm1) edge [w] (xt) 
	          (xt) edge [w] (xtp1)
	          (xtm1) edge [w] (otm1)
	          (xt) edge [w] (ot)
	          (xtp1) edge [w] (otp1)
	          (stm1) edge [a] (atm1)
	          (st) edge [a] (at)
	          (atm1) edge [w] (xt)
	          (at) edge [w] (xtp1)
}

\caption{Graphical representation of causal dependencies. Remember that $\mathbf{s}^*_t=(a_t,x_t,v_t)$. The solid lines correspond to the factors of the agent's generative model, $ p(\mathbf{s}_{t+1}|\mathbf{s}_t)$ and $ p(o_t,\mathbf{s}_t)$. The dashed lines correspond to the variational density $ q(\mathbf{s}_t|o_t,\mathbf{s}_{t-1})$. The dotted lines correspond to the environmental dynamics $  (x_{t+1}, v_{t+1}) = \mathbf{R}(x_t, v_t,a_{t+1})$ and the true generative densities $ p(o_t|\mathbf{s}_t^*)$. The wiggly line describes the dependency of the action $ p(a_{t+1}|\mathbf{s}_t)$ on the hidden states $ \mathbf{s}_t$ .}
\label{graphical_model}

\end{figure}

The causal structure of the complete model is shown in figure \ref{graphical_model}.

\subsubsection{Goal Directed Behavior}

If we just propagate and optimise our agent as it is now, it will look for a stable equilibrium with its environment and settle there. However, to be practically applicable to real-life problems, we have to instil some concrete goals in our agent. We can achieve this by defining states that it will expect to be in. Then the action states will try to fulfil these expectations.

In this concrete case, we want to propagate the agent for 30 timesteps and want it to be at $ x = 1.0$ for at least the last ten timesteps. As the agent's priors act on the hidden states, we introduce a hard-wired state which just represents the agents current position. We do this by explicitly encoding the agent's sense of position $o_{x,t}$ to the first dimension of the state vector $s_{1,t}$:
$$  q_\theta(s_{1,t} | \mathbf{s}_{t-1}, o_{x,t}, o_{h,t}, o_{a,t}) = \mathrm{N}(s_{1,t}; 0.1o_{x,t}, 0.01) $$

This can be seen as a homeostatic, i.e. vitally important, state parameter. E.g. the CO$_2$ concentration in our blood is extremely important and tightly controlled, as opposed to the possible brightness perceived at the individual receptors of our retina, which can vary by orders of magnitude. Though we might not directly change our behavior depending on visual stimuli, a slight increase in the CO$_2$ concentration of our blood and the concurring decrease in the pH will trigger chemoreceptors in the carotid and aortic bodies, which in turn will increase the activity of the respiratory centers in the medulla oblongata and the pons, leading to a fast and strong increase in ventilation, which might be accompanied by a subjective feeling of dyspnoea or respiratory distress. These hard-wired connection between vitally important body parameters and direct changes in perception and action might be very similar to our approach to encode the goal-relevant states explicitly.

But besides explicitly encoding relevant environmental parameters in the hidden states of the agent's latent model, we also have to specify the corresponding prior expectations (such as explicit boundaries for the pH of our blood). We do this by explicitly setting the prior over the first dimension of the state vector for $ t > 20$:
$$ p_\theta^t(s_{0,t}|\mathbf{s}_{t-1}) = \mathrm{N}(s_{0,t}; 0.1, 0.01), \mbox{if } t > 20 $$

While this kind of hard-coded inference dynamics and expectations might be fixed for individual agents of a class (i.e. species) within their lifetimes, these mappings can be optimised on a longer timescale over populations of agents by evolution. In fact, evolution might be seen as a very similar learning process, only on different spatial and temporal scales \citep{Watson2016, Baez2015, Harper2009}.

\subsubsection{The Free Energy Objective}

Now we have everything that we need to put our objective function together. We use the following form of the variational free energy bound (c.f section \ref{sec_free_energy}):
\begin{multline*}
F(o, \theta) = \left< - \ln p_\theta(o_{x,1}, ..., o_{x,T},o_{h,1}, ..., o_{h,T},o_{a,1}, ..., o_{a,t}|\mathbf{s}_1, ...., \mathbf{s}_T) \right>_q \\
 + D_{KL}(q_\theta(\mathbf{s}_1, ..., \mathbf{s}_T | o_{x,1}, ..., o_{x,T},o_{h,1}, ..., o_{h,T}, o_{a,1}, ..., o_{a,T}) || p_\theta(\mathbf{s}_1, \mathbf{s}_2, ..., \mathbf{s}_T) )
\end{multline*}
where $ \left<...\right>_q $ means the average with respect to the variational density 
$$  q_\theta(\mathbf{s}_1, ..., \mathbf{s}_T | o_{x,1}, ..., o_{x,T},o_{h,1}, ..., o_{h,T}, o_{a,1}, ..., o_{a,T}) $$

Using the above factorisation, the free energy becomes
\begin{multline*}
F(o, \theta) = \sum_{t=1}^T \bigg[ \left< -\ln p_\theta(o_{x,t}|\mathbf{s}_t) -\ln p_\theta(o_{h,t}|\mathbf{s}_t) -\ln p_\theta(o_{a,t}|\mathbf{s}_t) \right>_{q_\theta(\mathbf{s}_t|\mathbf{s}_{t-1},o_{x,t},o_{h,t},o_{a,t})} \bigg. \\
\bigg. + D_{KL} (q_\theta(\mathbf{s}_t|\mathbf{s}_{t-1},o_{x,t},o_{h,t},o_{a,t}) ||p_\theta  ( \mathbf{s}_t | \mathbf{s}_{t-1})) \bigg]
\end{multline*}

To evaluate this expression, we simulate several thousand processes in parallel, which allows us to approximate the variational density $ q$ just by a single sample $ \hat{\mathbf{s}}_t$ per process and per timestep, analogous to Stochastic Gradient Descent or the Variational Autoencoder, where only one sample per data point is enough, since the gradients depend on entire (mini-)batches of datapoints \citep{KingmaWelling2014, RezendeMohamedWierstra2014,Chung2015}. The sampling occurs following algorithm \ref{alg_free_energy}, where we use the closed form of the KL-Divergence for diagonal Gaussians:
\begin{multline*} 
D_{\mathrm{KL}}(q_\theta(\mathbf{s}_t | \hat{\mathbf{s}}_{t-1}, \hat{o}_{x,t}, \hat{o}_{h,t}, \hat{o}_{a,t}) || p_\theta(\mathbf{s}_t | \hat{\mathbf{s}}_{t-1})) =\\ D_{\mathrm{KL}} (\mathrm{N}(\mathbf{s}_t ;\boldsymbol\mu^q, \boldsymbol\sigma^q) || \mathrm{N}(\mathbf{s}_t ;\boldsymbol\mu^t, \boldsymbol\sigma^t))=
\sum_{i=1}^n \frac{1}{2} \left( 2 \ln \frac{\sigma_i^t}{\sigma_i^q} + \frac{(\sigma_i^q)^2 + (\mu_i^q - \mu_i^t)^2}{(\sigma_i^t)^2} - 1 \right)
\end{multline*}

\begin{algorithm}[t]
\caption{Sampling based approximation of the free energy cost. Concrete samples and values of a variable marked using a hat, e.g. $\hat{\mathbf{s}}_t$. $l$ is the sampling-based approximation of the expected likelihood of the observations under the variational density $q$, i.e. the accuracy term. $d$ is the KL-divergence between the variational density and the prior, i.e. the complexity term.}
\algnewcommand\algorithmicforeach{\textbf{for each}}
\algdef{S}[FOR]{ForEach}[1]{\algorithmicforeach\ #1\ \algorithmicdo}
\begin{algorithmic}
\State Initialize the free energy estimate with $F = 0$
\State Initialize $n_F$ agents with $\hat{\mathbf{s}}_0 = (0,...,0)$ and $(x_0,v_0) = (-0.5, 0.0)$.
\ForEach {Agent}
\For {$t = 1, ..., T$}
\State Sample an action $\hat{a}_{t}$ from $p_\theta(a_{t} | \hat{\mathbf{s}}_{t-1})$
\State Propagate the environment using $(x_t, v_t) = \mathbf{R}(x_{t-1}, v_{t-1}, \hat{a}_{t})$
\State Draw single observations $\hat{o}_{x,t}$ from $p_e(o_{x,t}|x_t)$
\State Set observation $\hat{o}_{h,t} = \exp(-\frac{(x_t-1.0)^2}{2\cdot 0.3^2})$ 
\State Set observation $\hat{o}_{a,t} = \hat{a}_{t}$ 
\State Draw a single sample $\hat{\mathbf{s}}_t$ from $q_\theta(\mathbf{s}_t | \hat{\mathbf{s}}_{t-1}, \hat{o}_{x,t}, \hat{o}_{h,t}, \hat{o}_{a,t})$
\State Calculate $l = -\sum_{i \in \{x,h,a\}} \ln p_\theta(\hat{o}_{i,t} | \hat{\mathbf{s}}_t)$
\State Calculate $d = D_{\mathrm{KL}}(q_\theta(\mathbf{s}_t | \hat{\mathbf{s}}_{t-1}, \hat{o}_{x,t}, \hat{o}_{h,t}, \hat{o}_{a,t}) || p_\theta(\mathbf{s}_t | \hat{\mathbf{s}}_{t-1}))$
\State Increment free energy $F = F + \frac{d + l}{n_p}$
\State Carry $\hat{\mathbf{s}}_t$ and $(x_t, v_t)$ over to the next timestep.
\EndFor
\EndFor
\\ \Return F
\end{algorithmic}
\label{alg_free_energy}
\end{algorithm}

The fact that we run a large number $n_F$ of processes in parallel allows us to resort to this simple sampling scheme for the individual processes.

The minimisation of the free energy with respect to the parameters $ \theta$ will improve the agents generative model $ p_\theta (o,\mathbf{s}) $, by lower bounding the evidence $ p_\theta(o)$ of the observations $ o$, given the generative model. Simultaneously it will make the variational density $ q_\theta (\mathbf{s}|o) $ a better approximation of the true posterior $ p_\theta(\mathbf{s}|o)$, as can be seen from the following, equivalent form of the free energy (c.f. section \ref{sec_free_energy}):

$
\begin{array}{rcl}
F(o, \theta) & = & <- \ln p_\theta(o|\mathbf{s})>_{q_\theta(\mathbf{s}|o)} \vphantom{ \left[ \left< -\ln p_\theta^{x,l}\right>_{q_\theta} \right]  }+ D_{KL}(q_\theta(\mathbf{s}|o)||p_\theta(\mathbf{s})) \\
& =\vphantom{ \left[ \left< -\ln p_\theta^{x,l}\right>_{q_\theta} \right]  } & - \ln p_\theta(o) + D_{KL}(q_\theta(\mathbf{s}|o)||p_\theta(\mathbf{s}|o))
\end{array}
$

Additionally, the parameters of the action function will be optimised, so that the agent seeks out expected states under its own model of the world, minimizing $ <- \ln p_\theta(o|\mathbf{s})>_{q_\theta(\mathbf{s}|o)} $.

\subsection{Optimization using Evolution Strategies}

Without action, the model and the objective would be very similar to the objective functions of \cite{KingmaWelling2014, RezendeMohamedWierstra2014, Chung2015}. So one could just do a gradient descent on the estimated free energy with respect to the parameters $ \theta$, using a gradient-based optimisation algorithm such as ADAM \citep{BaKingma2014}. However, to do this here, we would have to backpropagate the partial derivatives of the free energy with respect to the parameters through the dynamics of the world. I.e. our agent would have to know the equations of motions of its environment or at least their partial derivatives. As this is obviously not the case, and as many environments are not even differentiable, we have to resort to another approach.

It was recently shown that evolution strategies allow for efficient, large scale optimisation of complex, non-differentiable objective functions $ F(\theta)$ \citep{Salimans2017}. We apply this algorithm as follows: Instead of searching for a single optimal parameter set $ \theta^*$, we introduce a distribution on the space of parameters, which is called the "population density". We optimise its sufficient statistics to minimize the expected free energy under this distribution. The population density can be seen as a population of individual agents, whose average fitness is optimised, hence the name. The expected free energy over the population as function of the sufficient statistics $\psi$ of the population density $ p_\psi (\theta)$ is:

$$ \eta (\psi) = \left< F(\theta) \right>_{p_\psi (\theta)}$$

Now we can calculate the gradient of $ \eta$ with respect to the sufficient statistics $ \psi$:

$$
\nabla_\psi \eta(\psi) = \left< F(\theta) \nabla_\psi \ln p_\psi(\theta) \right>_{p_\psi (\theta)}
$$

Using a diagonal Gaussian for $ p_\psi(\theta) = \mathrm{N}(\theta; \mu^\psi, \sigma^\psi)$ the gradient with respect to the means $\mu^\psi$ is:
$$
\nabla_{\mu^\psi} \eta(\psi) = \left< F(\theta) \frac{1}{(\sigma^\psi)^2} \left( \theta - \mu^\psi \right) \right>_{p_\psi (\theta)}
$$

We will also optimise the standard deviations of our population density, using the corresponding gradient:

$$
\nabla_{\sigma^\psi} \eta(\psi) = \left< F(\theta) \frac{\left( \theta - \mu^\psi \right)^2 - \left(\sigma^\psi\right)^2}{(\sigma^\psi)^3} \right>_{p_\psi (\theta)}
$$

Drawing samples $ \epsilon_i$ from a standard normal distribution $ \mathrm{N}(\epsilon_i; 0,1)$, we can approximate samples from $ p_\psi(\theta)$ by $ \theta_i = \mu^\psi + \sigma^\psi \epsilon_i$. Thus, we can approximate the gradients via sampling by:

\begin{equation}
\nabla_{\mu^\psi} \eta(\psi) \approx \frac{1}{n_\textrm{pop}} \sum_{i=1}^{n_\textrm{pop}} F(\theta_i) \frac{\epsilon_i}{\sigma^\psi}
\end{equation}

and

$$
\nabla_{\sigma^\psi} \eta(\psi) \approx \frac{1}{n_\textrm{pop}} \sum_{i=1}^{n_\textrm{pop}} F(\theta_i) \frac{\epsilon_i^2 - 1}{\sigma^\psi}
$$

For reasons of stability we are not optimising $ \sigma^\psi$ directly, but calculate the standard deviations using:

$$ \sigma^\psi = \mathrm{softplus}(\tilde{\sigma}^\psi) + \sigma^\psi_\mathrm{min}$$

with $ \mathrm{softplus}(x) = \ln(1+e^x)$. By choosing $ \sigma^\psi_\mathrm{min} = 10^{-6}$ constant and optimising $ \tilde{\sigma}^\psi$ we prevent divisions-by-zero and make sure that there is no sign-switch during the optimisation. The chain rule gives:

\begin{equation}
\nabla_{\tilde{\sigma}^\psi} \eta(\psi) \approx \frac{1}{n_\textrm{pop}} \sum_{i=1}^{n_\textrm{pop}} F(\theta_i) \frac{\epsilon_i^2 - 1}{\sigma^\psi}\frac{\partial \sigma^\psi}{\partial \tilde{\sigma}^\psi} = \frac{1}{n_\textrm{pop}} \sum_{i=1}^{n_\textrm{pop}} F(\theta_i) \frac{\epsilon_i^2 - 1}{\sigma^\psi}\frac{\exp \left( \tilde{\sigma}^\psi \right)}{1 + \exp \left( \tilde{\sigma}^\psi \right)}
\end{equation}

Using these gradient estimates, we now optimise the expected free energy bound under the population density using ADAM as gradient based optimiser \citep{BaKingma2014}. The corresponding pseudocode is shown in algorithm~\ref{alg_optimisation}. 

\begin{algorithm}[t]
\caption{Optimisation of the free energy bound.}
\algnewcommand\algorithmicforeach{\textbf{for each}}
\algdef{S}[FOR]{ForEach}[1]{\algorithmicforeach\ #1\ \algorithmicdo}
\begin{algorithmic}
\State Initialize the population density on the parameters $\theta$ using randomised sufficient statistics $\psi =  \{\mu^\psi, \tilde{\sigma}^\psi\}$
\While {Expected free energy $\eta(\psi)$ has not converged}
\State Draw $n_\textrm{pop}$ samples $ \theta_i = \mu^\psi + \sigma^\psi \epsilon_i$, $\epsilon_i \sim \mathrm{N}(0,1)$
\ForEach {Sample}
\State Approximate the free energy bound $F(\theta_i)$ using $n_F$ processes
\EndFor
\State Approximate the gradients $\nabla_\psi \eta$ using equations~1 and 2.
\State Perform a parameter update on $\psi$ using ADAM with these gradient estimates.
\EndWhile
\end{algorithmic}
\label{alg_optimisation}
\end{algorithm}

\subsection{Constrained Sampling from the Learned Generative Model}

\label{sec_constrained_sampling}

Once the agent has been optimised, we can not only propagate it through the environment or draw unconstrained samples from its generative model of the world, we can also use a Markov-Chain Monte Carlo algorithm to draw constrained samples from its generative model \citep{RezendeMohamedWierstra2014}. This can on one hand be used to impute missing inputs. E.g. by sampling from

 $$p_\theta( o_{h,1}, o_{h,2}, ..., o_{h,T} | o_{x,1}, o_{x,2}, ..., o_{x,T}, o_{a,1}, o_{a,2}, ..., o_{a,T})$$ 
 
we can get the agents estimate on the homeostatic sensory channel, given the proprioceptive and spatial sensory channels. By using multiple samples we can even get sensible bounds on the uncertainty of these estimates by approximating the full distribution.
 
 On the other hand, we can ask the agent about its beliefs about the course of the world, given its actions (in terms of its proprioceptive channel $o_a$). Thus, we can get an explicit readout about the agents beliefs about its influence on the world.
 
 $$p_\theta( o_{x,1}, o_{x,2}, ..., o_{x,T}, o_{h,1}, o_{h,2}, ..., o_{h,T} | o_{a,1}, o_{a,2}, ..., o_{a,T})$$ 
 
The required algorithm is developed in appendix F of \cite{RezendeMohamedWierstra2014} and described here as algorithm~\ref{alg_constrained_sampling}. The basic idea is to use the de-noising properties of autoencoders due to the learned, abstract and robust representation and their ability to generate low-dimensional representations capturing the regularity and systematic dependencies within the observational data. Thus, the workings of this algorithm can be understood as follows: First, all but the given sensory channels are randomly initialised. These partly random sensory observations are now encoded using the variational distribution $q$. The resulting state tries to represent the observation within the low-dimensional, robust representation learned by the agent and should thereby be able to remove some of the "noise" from the randomly initialised channels, just in line with the classic idea of an autoencoder \citep{Hinton2006}. From this variational distribution a sample is drawn, which can be used to generate new, sensory samples, that are already a bit less noisy. Now the known observations can be reset to their respective values and the denoised observations can again be encoded, using the variational density $q$. By iteratively encoding the denoised samples together with the given sensory inputs, the iterative samples from the abstract, robust representation will converge to the most probable cause (in terms of the hidden states) for the actually observed sensations under the generative model. As the variational density and the generative model capture the regularities and dependencies within the observations, the observations generated from this representation will converge to the distribution of the unknown observations, given the observed channels. \cite{RezendeMohamedWierstra2014}  provided a proof that this is true, given that the unobserved channels are not initialised too far away from the actual values.
 
\begin{algorithm}[t]
\caption{Markov-Chain Monte Carlo algorithm for constrained sampling from the agent's generative model. Concrete samples and values of a variable marked using a hat, e.g. $\hat{\mathbf{s}}_t$.}
\algnewcommand\algorithmicforeach{\textbf{for each}}
\algdef{S}[FOR]{ForEach}[1]{\algorithmicforeach\ #1\ \algorithmicdo}
\begin{algorithmic}
\State \textbf{Given:} Proprioceptive sensory inputs $o_{a,1},...,o_{a,T}$ 
\State Initialize $n_s$ processes with $\hat{\mathbf{s}}_0 = (0,...,0)$
\ForEach {Process}
\For {$t = 1, ..., T$}
\State Initialize $\hat{o}^0_{x,t}$ with sample from $\mathrm{N}(o^0_{x,t};0, 0.01)$
\State Initialize $\hat{o}^0_{h,t}$ with sample from $\mathrm{N}(o^0_{h,t};0, 0.01)$
\For {$i = 1, ..., n_i$}
\State Sample $\hat{\mathbf{s}}^i_t$ from $q_\theta(\mathbf{s}^i_t | \hat{\mathbf{s}}_{t-1}, \hat{o}^{i-1}_{x,t}, \hat{o}^{i-1}_{h,t}, o_{a,t})$
\State Sample new estimates $ \hat{o}^i_{x/h,t}$ from each likelihood $ p_\theta(o^i_{x/h,t} | \hat{\mathbf{s}}^i_t)$
\EndFor
\State Set $\hat{\mathbf{s}}_t = \hat{\mathbf{s}}^{n_i}_t$
\State Set $\hat{o}_{x,t} = \hat{o}^{n_i}_{x,t}$
\State Set $\hat{o}_{h,t} = \hat{o}^{n_i}_{h,t}$
\State Return $\hat{o}_{h,t}$ and $\hat{o}_{a,t}$
\State Carry $\hat{\mathbf{s}}_t$ over to the next timestep.
\EndFor
\EndFor
\end{algorithmic}
\label{alg_constrained_sampling}
\end{algorithm}

\subsection{Experimental Set Up and Parameters}

The experiments were performed on a desktop PC equipped with a 2013 NVIDIA GTX Titan GPU. The parameter values used in the optimisation algorithm~\ref{alg_optimisation} and the required estimation of the free energy bound using algorithm~\ref{alg_free_energy} are shown in table \ref{tab_params}. Note that we approximate the free energy by a single process ($n_F = 1$) for each sample from the population density. This is possible, as we draw many ($n_\textrm{pop} = 10^4$) samples from the population density. Using more processes for each sample, while keeping the total number $n_\textrm{tot} = n_\textrm{pop}n_F$ of simulated processes constant, results in a worse convergence behavior, as the coverage of the parameter space has to be reduced, while the total variability due to the stochastic approximation of the total bound stays approximately constant. The full code of this implementation and the scripts to reproduce all figures in this paper can be downloaded here: \url{https://www.github.com/kaiu85/deepAI_paper}. The code is written in Python 2.7, using Theano \citep{Theano2016} for GPU optimised Tensor-Operations.

\section{Results}

\begin{figure}
\makebox[\textwidth][c]{\includegraphics[width=1.0\textwidth]{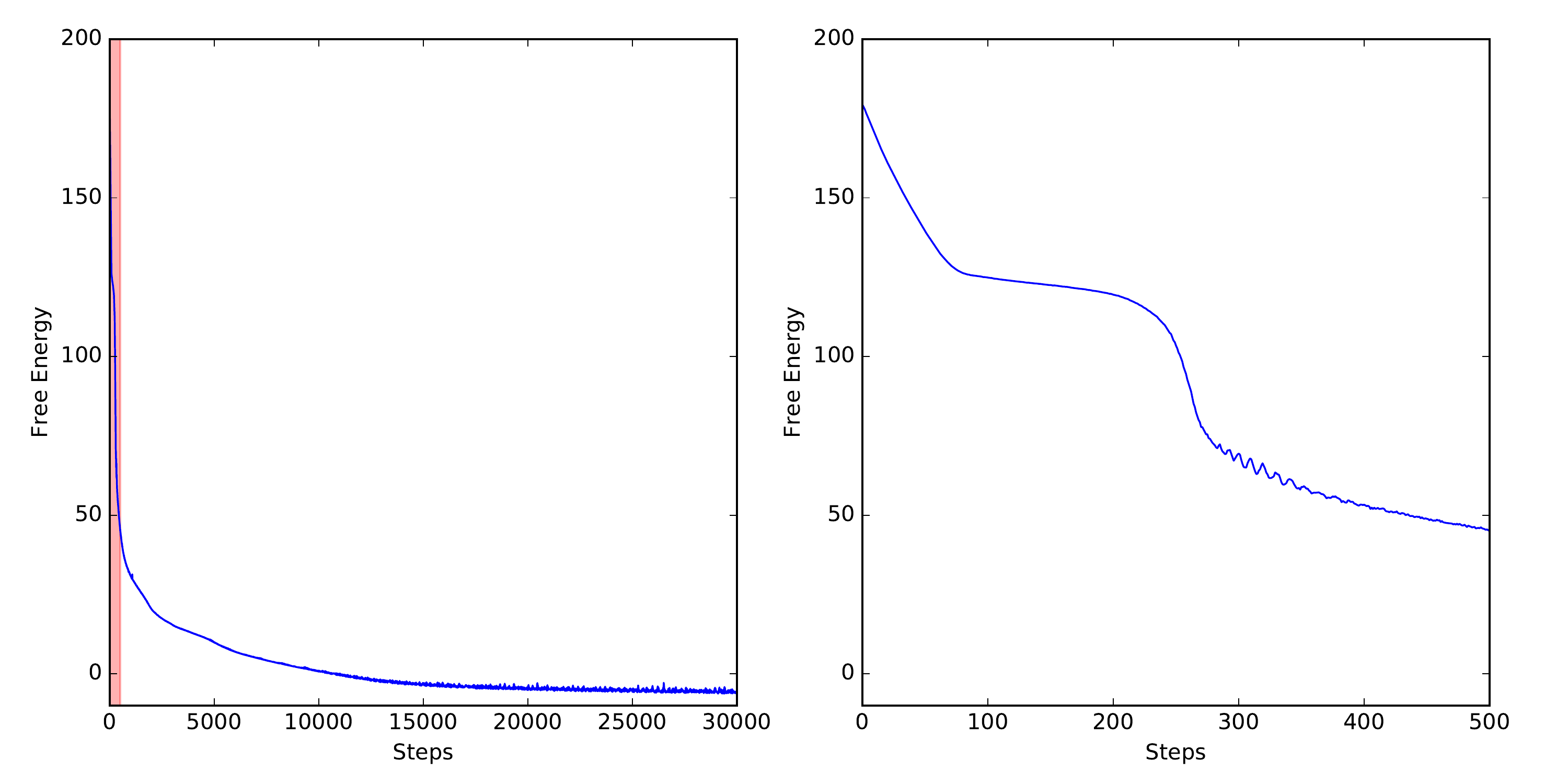}}
\caption{Convergence of an active inference agent, using parameters in table \ref{tab_params}. The area shaded in red in the left plot was enlarged in the right plot. }
\label{fig_convergence}
\end{figure}

\begin{figure}
\makebox[\textwidth][c]{\includegraphics[width=1.0\textwidth]{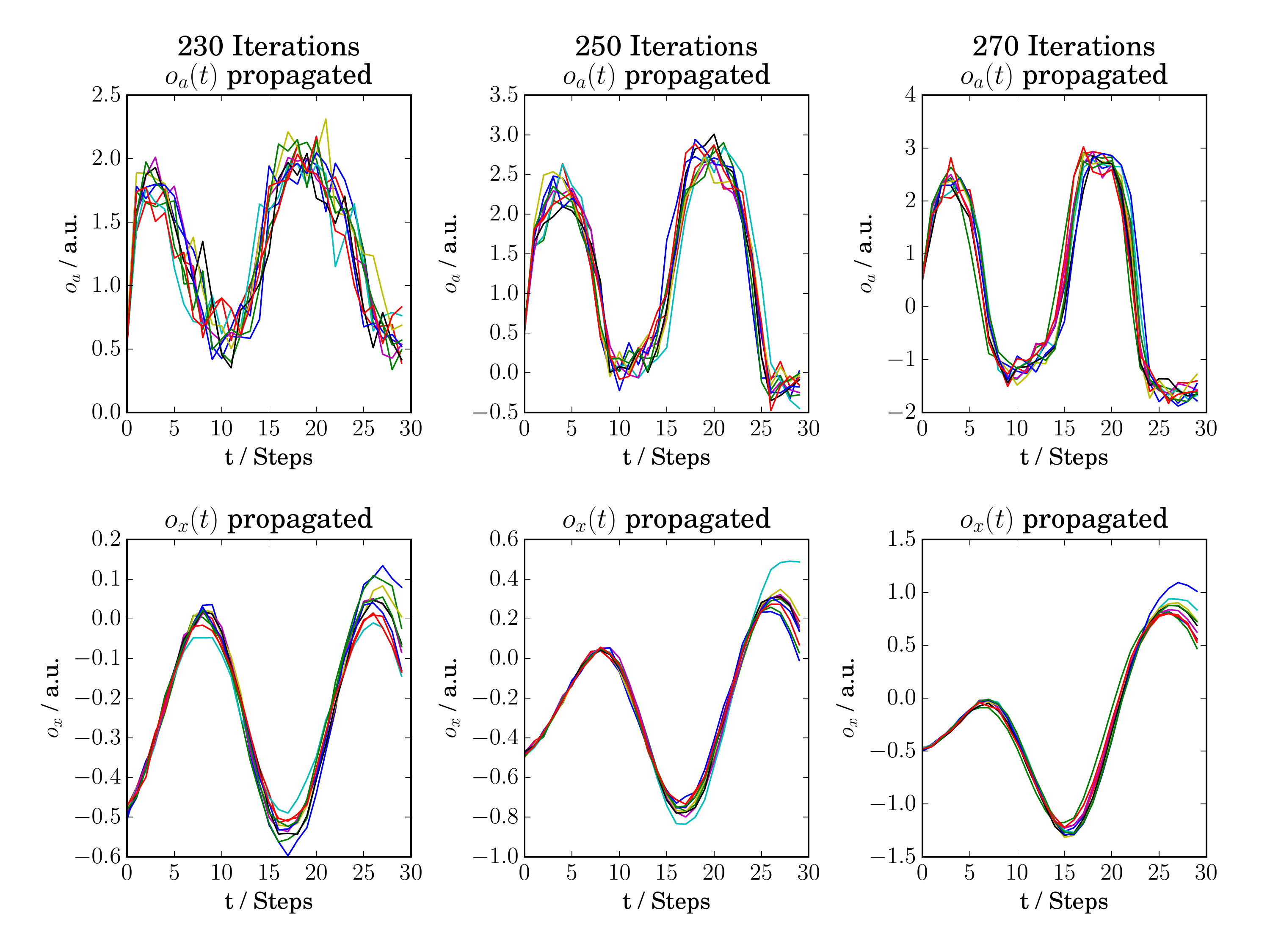}}
\caption{ The agent's performance after 230, 250, and 270 training steps, using the mean parameters of the population density. It has just realised that by moving uphill a bit to the left (from $ t = 15$ to $t = 20$), it can reach a higher position (around $t = 27$) than by directly going upwards (c.f. $t = 9$). Shown are the agent's proprioception $o_a$ (upper row), and its sense of position $ o_x$ (lower row). }
\label{fig_insight}
\end{figure}

The evolution strategies based optimisation procedure used less than 300 MB of GPU memory and took less than 0.4 s per iteration. Figure \ref{fig_convergence} shows the convergence of the free energy bound within 30,000 training steps (less than 3.5 hours). It quickly converges from its random starting parameters to a plateau, on which it tries to directly climb the hill and gets stuck at the steep slope. However, after only about 250 updates of the population density it discovers, that it can get higher by first moving in the opposite direction, thereby gaining some momentum, which it can use to overcome the steep parts of the slope . This insight leads to a sudden, rapid decline in free energy \citep{Friston2017}. The rapid development of the agent's strategy and its quick adoption of the insight, that an initial movement away from its target position is beneficial, is illustrated in figure~\ref{fig_insight}.

\begin{table}
\begin{tabular}{rcc}
\hline
\textbf{parameter} & \textbf{description} & \textbf{value} \\
\hline
\hline
$\alpha$ & learning rate of ADAM optimiser & 0.001 \\
\hline
$(\beta_1,\beta_2)$ & \makecell{exponential decay rates for moment\\ estimation of ADAM optimiser} & (0.9, 0.999) \\
\hline
$\epsilon$ & \makecell{noise parameter\\ of ADAM optimiser} & $10^{-8}$ \\
\hline
$n_F$ & \makecell{number of processes to\\ approximate free energy for each\\ sample from the population density} & $ 1$ \\
\hline
$n_\textrm{pop}$ & \makecell{number of samples from\\ the population density} & $10^{4}$ \\
\hline
\end{tabular}
\caption{Parameter values used in algorithms \ref{alg_free_energy} and \ref{alg_optimisation}.}
\label{tab_params}
\end{table}

\begin{figure}
\makebox[\textwidth][c]{\includegraphics[width=1.0\textwidth]{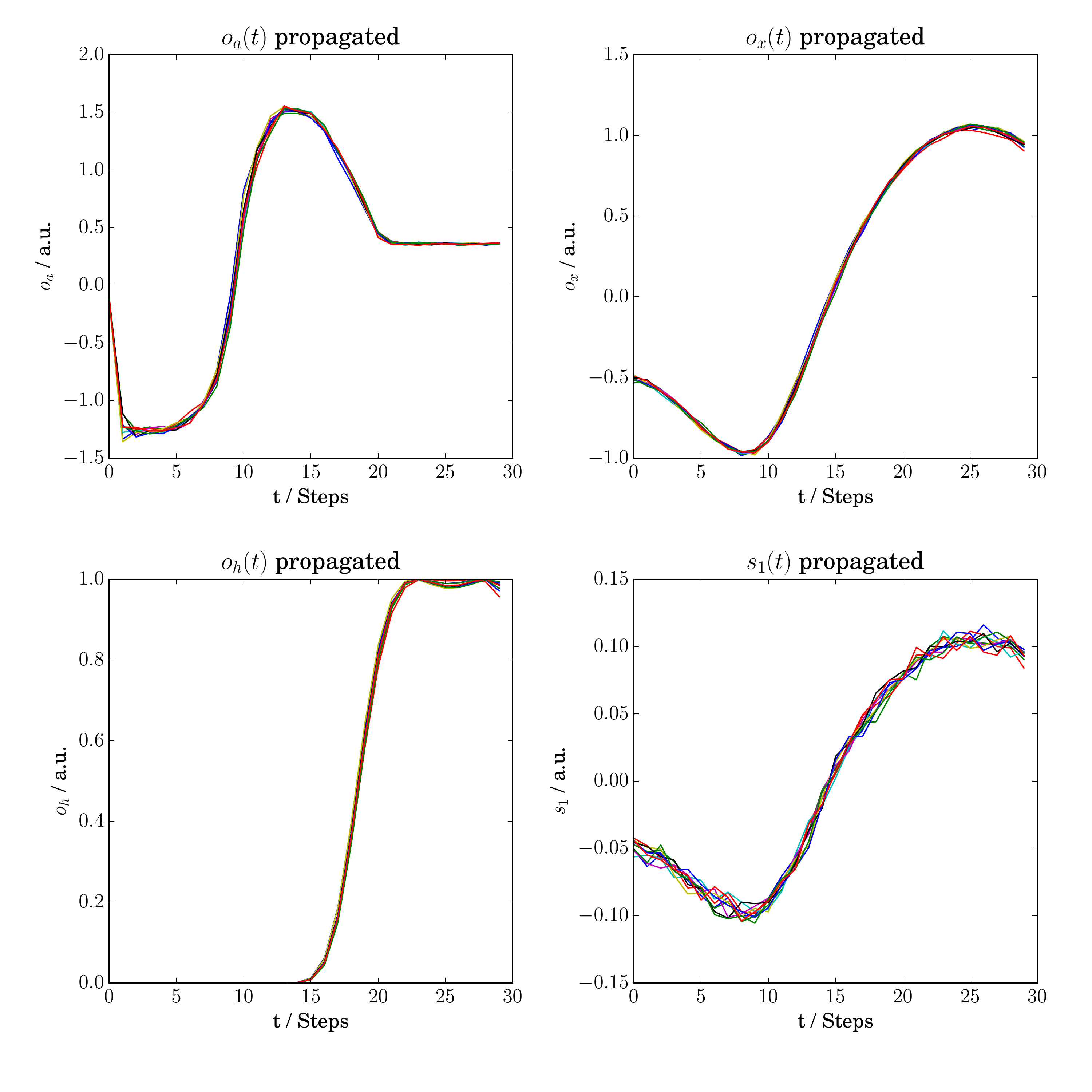}}
\caption{ The agent's performance after 30,000 training steps, using the mean parameters of the population density. It tightly sticks to a strategy, shown here in terms of the proprioceptive sensory channel $ o_a$ (upper left). The resulting trajectory (shown in terms of $ o_x$, upper right) first leads uphill to the left, away from the target position, to gain momentum and overcome the steep slope at $ x = 0$. The nonlinearly modified sensory channel $ o_h$ is shown on the lower left and the "homeostatic" hidden state $ s_1$ on the lower right. }
\label{fig_propagated}
\end{figure}

The agent's trajectory after about 30,000 training steps is shown in figure~\ref{fig_propagated}: It takes a short left swing to gain the required momentum to overcome the steep slope at $x = 0$, then directly swings up to its target position $x = 1.0$, and stays there by applying just the right force to counteract the non-zero downhill force at the target position $x = 1.0$.

\begin{figure}
\makebox[\textwidth][c]{\includegraphics[width=1.0\textwidth]{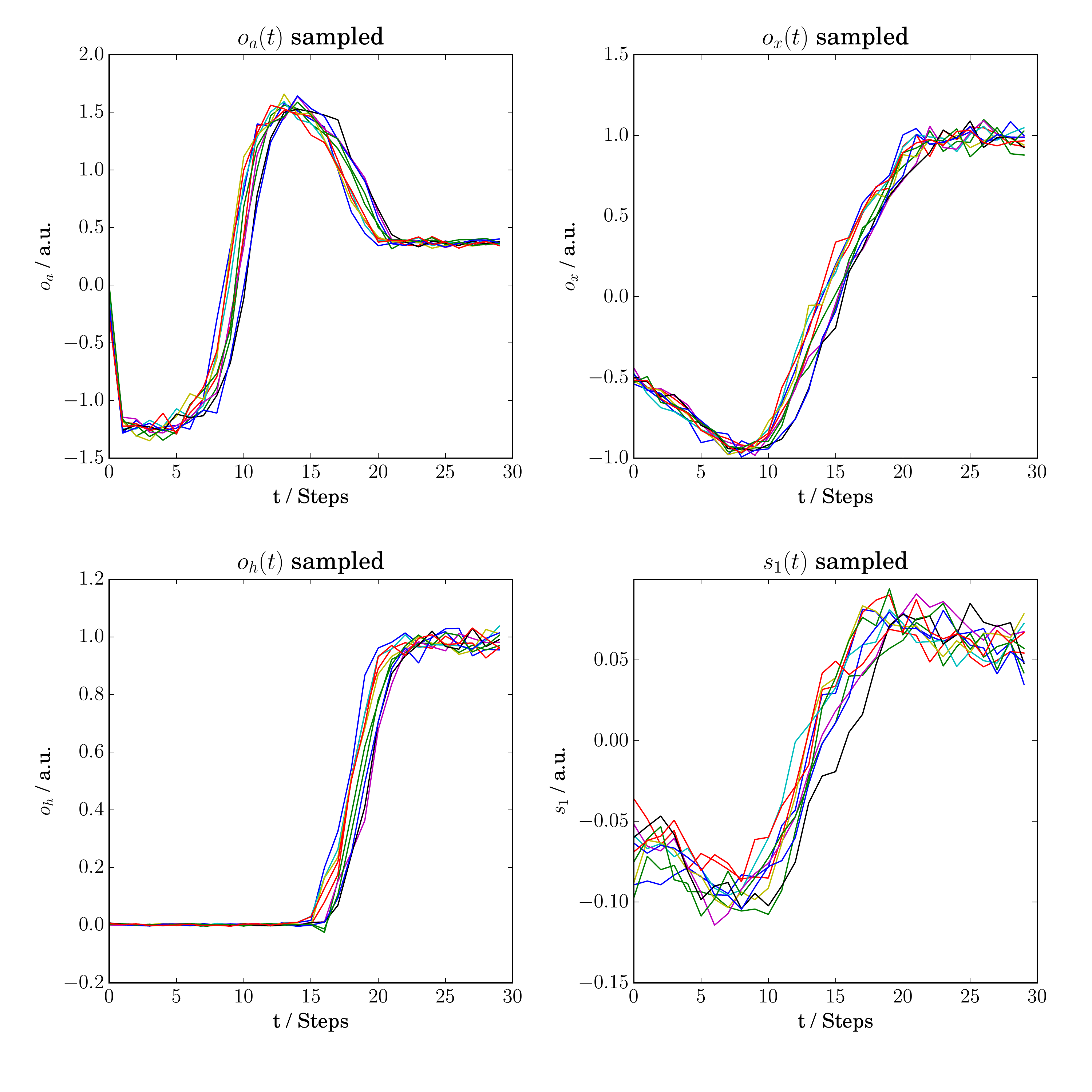}}
\caption{ Ten processes sampled from the agent's generative model of the world after 30,000 training steps, using the mean parameters of the population density. Shown are the prior expectations on the proprioceptive channel $ o_a$ (upper left), the agent's sense of position $ o_x$ (upper right), a nonlinear transformation of the position $ o_h$, and the agent's prior expectation on its "homeostatic" state variable $ s_1$. Note that each distribution is approximated by a single sample per timestep per process.}
\label{fig_sampled}
\end{figure}

After 30,000 optimisation steps, the agent has also developed quite some understanding of its environment. We can compare figure~\ref{fig_propagated}, which was generated by actually propagating the agent in its environment, with figure~\ref{fig_sampled}, which was generated by sampling from the agents generative model of its environment, \textit{without any interaction} with the environment. We see that the agent did not only learn the timecourse of its proprioceptive sensory channel $ o_a$ and its sense of position $ o_x$, but also the - in this setting irrelevant - channel $ o_h$, which is just a very nonlinear transformation of its position. Note that each panel of figure~\ref{fig_sampled} shows 10 processes sampled from the generative model as described in algorithm~\ref{sampling_from_gm}. Note that we are approximating each density just by a single sample per timestep and per process. Thus, although our estimates seem a bit noisy, they are very consistent with the actual behavior of the agent and the variability can easily be reduced by averaging over several processes.

\begin{figure}
\makebox[\textwidth][c]{\includegraphics[width=1.0\textwidth]{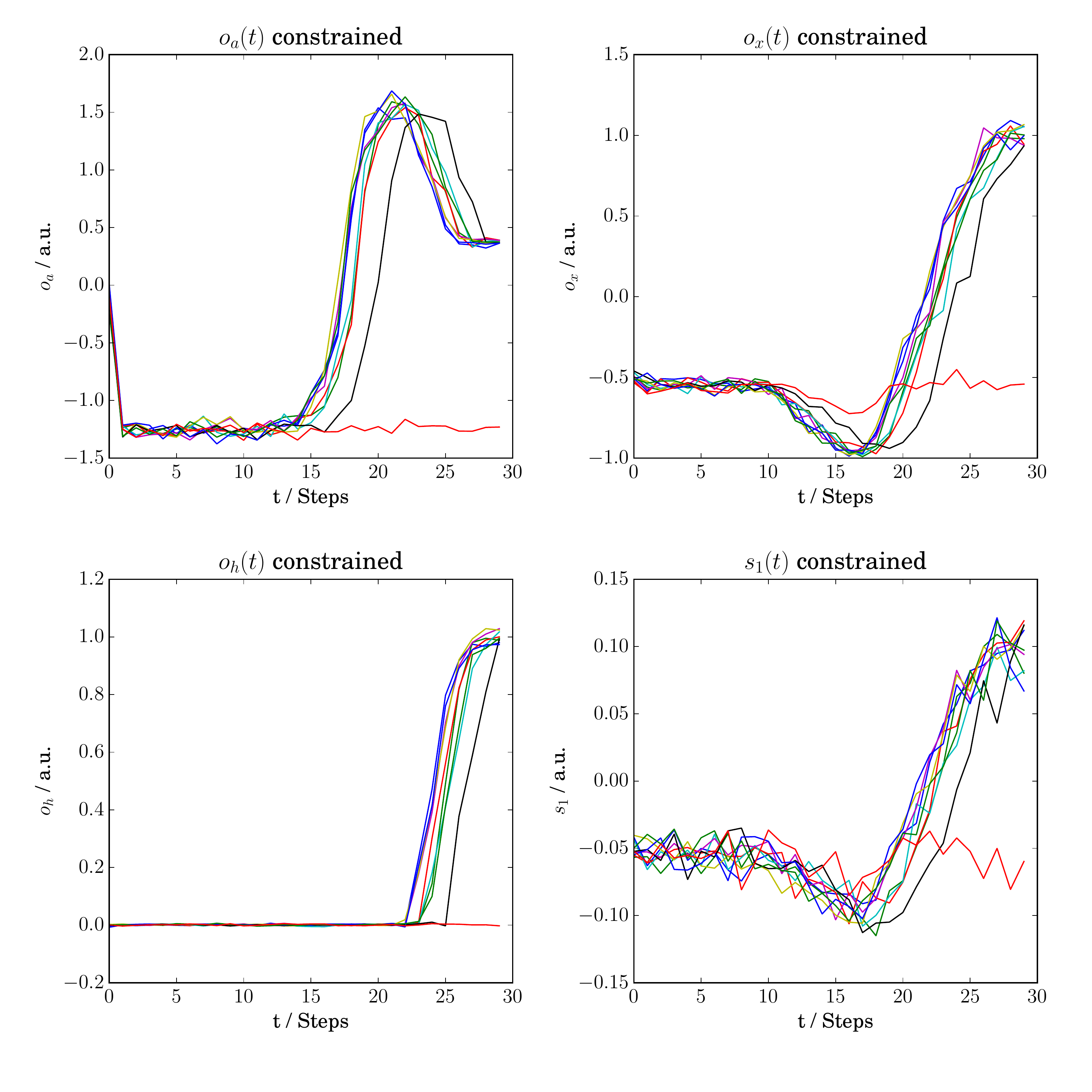}}
\caption{ Ten processes sampled from the agent's generative model of the world after 30,000 training steps, constrained on a given trajectory of the proprioceptive input $ o_a$ (upper left), using the mean parameters of the population density. Shown are the constrained expectations on the proprioceptive channel $ o_a$ (upper left), the agent's sense of position $ o_x$ (upper right), a nonlinear transformation of the position $ o_h$, and the agent's constrained expectation on its "homeostatic" state variable $ s_1$. Note that each distribution is approximated by a single sample per timestep per process.}
\label{fig_constrained}
\end{figure}

Having learned a generative model of the environment, we can not only propagate it freely, but we can also use it to test beliefs of the agent, given some a priori assumptions on the timecourse of certain states or sensory channels, using the algorithm described in section~\ref{sec_constrained_sampling}. We sampled the agent's prior beliefs about his trajectory $ o_{x,t}, o_{h,t}$, given its proprioceptive inputs $  o_{a,t}$, i.e. $ p(o_{x,1},...,o_{x,T},o_{h,1},...,o_{h,T}|o_{a,1},...,o_{a,T}) $ . Using the above example we took the average timecourse of the proprioceptive channel for the true interaction with the environment and shifted it back 10 timesteps. The results are shown in figure~\ref{fig_constrained}. First, we see that not all of the 10 sampled processes did converge. This might be due to the Markov-Chain-Monte-Carlo-Sampling approach, in which the chain has to be initialised close enough to the solution to guarantee convergence. However, for 9 out of 10 processes, the results look very similar to the true propagation (figure~\ref{fig_propagated}) and the unconstrained samples from the generative model (figure~\ref{fig_sampled}), only shifted back 10 timesteps.

\section{Discussion and Outlook}

\subsection{A Scalable and Flexible Active Inference Implementation} In this paper we have shown that the free energy objective proposed by the active inference framework in cognitive neuroscience \citep{FristonAl2010} allows an agent to find a solution to the mountain car environment, while concurrently building a generative model of itself and its environment. By implementing the internal dynamics of our agent using deep neural networks \citep{Lecun2015, Goodfellow2016} and recurrent dynamics \citep{Karpathy2015}, it is able to approximate the true generative process of its environment by its own generative model. Furthermore, as we are using an efficient, black-box optimiser for non-differentiable objective functions, the agent does not require direct access to any information regarding on true generative process or its derivates for any given environment. As the implementation and optimisation of this agent uses methods that are applied to complex, large scale problems in machine learning and artificial intelligence \citep{Chung2015, Rezende2016, KingmaIAF}, we hope that this class of agent can be of further use to demonstrate that active inference is able to solve more complex and realistic problems, such as the Atari and 3D robotic environments from OpenAI Gym \citep{OpenAIGym}. The Atari  environments require an agent to learn to play Atari games from raw pixel or RAM input, while the 3D robotic environments use the MujoCo physics engine \citep{Todorov2012}, to accurately simulate 3D robotic control problems. 

\subsection{Comparison to Original Implementation} In contrast to the original implementation of \cite{FristonAl2010}, our implementation is formulated in discrete timesteps without an explicit representation of generalised coordinates. I.e. our agent has to learn how its observations of the position $o_x$ are related between successive timesteps and to form its own representation of its velocity. Furthermore, the agent's generative model of the world has - in contrast to the former work - not the same functional form as the true generative process. I.e. our agent also has to learn an approximation of the true dynamics in terms of its own generative model. This is possible due to the implementation of the agents generative model in terms of a high-dimensional, recurrent neural network. These structures were shown to be able to efficiently learn and represent the underlying structure in complex time-series data from real-life applications \citep{Karpathy2015, Hinton2015, Chung2015}. Our agent does also have no access to the partial derivatives of its sensory inputs with respect to its actions, as these also depend on a knowledge of the true generative process, which real agents usually do not possess. By using evolution strategies \citep{Salimans2017}, we are able to derive stochastic estimates of these gradients that enable us to train a full active inference agent despite these constraints. Furthermore, as the structure of our agent and the utilised transfer functions and optimisers are directly adopted from large-scale machine learning \citep{Chung2015}, we hope that our agent will scale to more realistic and rich environments, showcasing the potential of active inference in terms of rich emergent behavior and the simultaneous optimisation of a generative model of the environment.

\subsection{Comparison to a recent "Action-Oriented" Implementation of Active Inference} Another implementation of Active Inference was recently introduced by \cite{Baltieri2017}. Similar to our approach, the authors do not give their agent access to the generative process, neither in terms of the functional form of the agent's generative model, nor by providing it with the partial derivatives of its full sensory inputs with respect to its actions. However, their work differs in several crucial aspects from ours: Their agent is formulated in continuous time, using partial differential equations. These equations are simple enough to explicitly discuss their dynamics. In contrast, we use a discrete time model with high-dimensional and highly nonlinear transfer functions, precluding us from a classical analytical treatment of the resulting dynamics. However, the very flexible form of the generative model in our approach allows our agent to learn an almost perfect approximation of the true dynamics, as shown in figure~\ref{fig_sampled}. Furthermore, they circumvent the lack of explicit partial derivatives of the sensory inputs with respect to the agent's actions, by subdividing its sensory inputs into exteroceptive and proprioceptive channels, where they use action to only suppress prediction errors in the proprioceptive channels. They call this an "Action-Oriented" approach. In terms of our model this would be equivalent to ignoring the partial derivatives of the "exteroceptive" sensory channels $o_x$ and $o_h$ with respect to the action $a$ and updating the parameters of the action function only using the gradients of the expected sensory surprise $<-\ln p_\theta(o_{a,t}|\mathbf{s}_t)>_{q(\mathbf{s}_t|\mathbf{s}_{t-1},o_{x,t},o_{h,t},o_{a,t})}$ under the population density. On one hand, this leads to an arbitrary subdivision of sensory channels and necessitates the proprioceptive channel $o_a$, which in our approach is not really necessary to successfully build a generative model of the environment and reach the goals defined by the agent's prior expectations (c.f. supplementary figures 9-11). On the other hand, this might severely hinder or at least delay the agent's learning in more complex and realistic environments (c.f. supplementary figures 12-14). Third, the full active inference framework \citep{FristonKilnerHarrison2006, FristonAl2010} underlines the crucial role of those partial derivatives and hints at possible approximations implemented in the brain. For example the retinotopic maps in the early visual system, e.g. in the superficial and deep layers of the superior colliculus, allow a quick and simple inversion of sensory input with respect to small eye movements \citep{FristonAl2010}. Note that in active inference, action selection per se is generally restricted to the sorts of inputs that the action system has access to (i.e., proprioceptive inputs). However, learning the parameters of the action mapping – as in this work – is clearly accountable to all sorts of inputs. This is effectively what we observe in our setup (c.f. supplementary figures 12-14).

\subsection{Action Function as State-Action Policy} The fixed mapping from the approximate posterior on states of the world, given the agents observations, to a distribution on possible actions is known as a state-action policy in control theory. It presupposes a Bayes optimal action for every state of the world. These are a very common forms of policies; however, they are not universal policies in the sense that many actions depend upon previously encountered states. However, as we allow the agent to develop its own, very flexible representation of the current state of the world, it can basically include a compressed representation of the history of previous states, if this information might be required to guide its action. Indeed, our agent shows this behavior, as it developed a representation of its current velocity, which can be regarded as difference between its current and its previous position. Otherwise it would not have been able to successfully solve the mountain car problem, as the agent had to learn to move left from its initial position, to acquire some additional momentum, while it later had to accelerate to the right at the exact same position, but after it had acquired the required additional momentum, which  allowed it to climb the steep slope at $x = 0.0$.

\subsection{Emergent Curiosity} Optimising its generative model of the world gives the agent a sense of epistemic value \citep{FristonEpistemicValue}. I.e. it will not only seek out the states, which it expects, but it will also try to improve its general understanding of the world. This is similar to recent work on artificial curiosity \citep{Pathak2017}. However, in contrast to this work, our agent is not only interested in those environmental dynamics which it can directly influence by its own action, but also tries to learn the statistics of its environment which are beyond its control.

\subsection{Constrained Sampling for Understanding the Learned Models} We were able to demonstrate constrained sampling from the agent's generative model of the world. Comparing the constrained samples (figure~\ref{fig_constrained}) to the actual interaction (figure~\ref{fig_propagated}) with the environment and the unconstrained samples (figure~\ref{fig_sampled}), they look reasonable, but deviate from the true dynamics that one would expect given the conditioned time course: We constrained the sampling on the mean timecourse of $o_a$ from the agent's true interaction with the world, shifted back in time by 10 time steps. This lead to samples of the other variables, which were also shifted back in time. However, during the first 10 time steps the agent believes that it would stay at its initial position, despite its strong push to the left. This might be due to the fact that the relative weighting of the agent's prior expectations on its position is quite strong. Accordingly, it tightly sticks to the optimal trajectory, as soon as it has learned it. Thus, the dynamics it infers deviate from the true dynamics, as it only ever experiences a very small subset of its phase space. However, if you put the agent in a noisy environment with larger stochastic fluctuations, where it is forced to explore and encounter a wider variety of dynamic states, it should learn a more complete and realistic model of the environmental dynamics. This hope also rests on the fact that by taking away its effector organs, our agent can be reduced to a generative recurrent latent variable model. This class of models has been able to model and generate very complex time series data, such as spoken language on the basis of its raw audio waveform \citep{Chung2015}. Thinking about autonomous agents in a real environment (i.e. robots or autonomous vehicles), constrained sampling from an agent's generative model might be a good way to understand its beliefs about how the world might react to given actions or events, opening up the "black-box" associated with classical deep reinforcement learning algorithms. 

\subsection{Benefits of the Evolution Strategies Optimiser} Evolution strategies allow for exploration of the environment without requiring the agent's action functions to be probabilistic: Using standard reinforcement-learning algorithms, such as deep Q-learning \citep{Mnih2015}, the only way that the agent can discover new states, which do not correspond to its current, local optimum, is by requiring it to stochastically take actions from time to time. This might be by sampling completely random actions or lower bounding the standard-deviation of its actions. Here, no such artificially forced exploration is required, as the evolution strategies algorithm explores new solutions on the parameter space \citep{Salimans2017}. So our agent can (and actually does) settle to an (almost) fully deterministic policy, if such a policy exists in the respective environment. Moreover, as the gradient estimates only depend on the full free energy, after the completion of each individual simulation, they are also resilient against long time horizons and sparse rewards \citep{Salimans2017}. 

\subsection{Variational Bayesian Perspective on Evolution Strategies} Although we are using evolution strategies mainly as a black-box optimiser for our non-differentiable objective function, namely the sampling based approximation of the free energy bound on the agent's surprise, a subtle but important reinterpretation of the population density highlights the ubiquitous role of variational Bayes and the following minimisation of variational free energy. In brief, if we rename the population density $p_\psi(\theta)$ with $q_\psi(\theta)$ and write the likelihood as $p(o|s,\theta)$ instead of $p_\theta(o|s)$ and the variational density as $q(s|o,\theta)$ instead of $q_\theta(s|o)$, then the expected free energy becomes:
$$\eta(\psi,o) = \left< \left<-\ln p(o|s,\theta)\right>_{q(s|o,\theta)} + D_\textrm{KL}(q(s|o,\theta)||p(s|\theta))\right>_{q_\psi(\theta)}$$
If we now add a complexity penalty $D_\textrm{KL}(q_\psi(\theta)||p(\theta))$ on the population density, based on a prior $p(\theta)$ on the parameters $\theta$, the sum of the expected variational free energy under the population density plus this KL divergence becomes
\begin{multline*}
\eta(\psi,o) + D_\textrm{KL}(q_\psi(\theta)||p(\theta)) \\ 
= \left< \left<-\ln p(o|s,\theta)\right>_{q(s|o,\theta)} + D_\textrm{KL}(q(s|o,\theta)||p(s|\theta))\right>_{q_\psi(\theta)} + \left< \ln q_\psi(\theta) - \ln p(\theta)\right>_{q_\psi(\theta)} \\
=  \left< \left<-\ln p(o|s,\theta)\right>_{q(s|o,\theta)} + \left< \ln q(s|o,\theta) - \ln p(s|\theta) \right>_{q(s|o,\theta)} + \ln q_\psi(\theta) - \ln p(\theta)\right>_{q_\psi(\theta)} \\
= \left< -\ln p(o|s,\theta) + \ln q(s|o,\theta) - \ln p(s|\theta) + \ln q_\psi(\theta) - \ln p(\theta))\right>_{q(s|o,\theta)q_\psi(\theta)} \\
= \left< -\ln p(o|s,\theta) + \ln q(s|o,\theta) q_\psi(\theta) - \ln  p(s|\theta) p(\theta))\right>_{q(s|o,\theta)q_\psi(\theta)}
\end{multline*}
This is just the variational free energy under a proposal density $q_\psi(s,\theta | o) = q(s|o,\theta)q_\psi (\theta)$ and prior $p(s,\theta) = p(s|\theta)p(\theta)$ that cover both states and parameters, i.e. full variational Bayes (c.f. \citep{KingmaWelling2014}, appendix F).
On this view, the population dynamics afford a general and robust, ensemble based scheme for optimising model parameters with respect to the variational free energy of beliefs over both model parameters and latent states, neglecting prior information (or using a noninformative prior) on the parameters. This is similar to the more general formulation of \cite{Friston2008}, who also absorbed deep inference and deep learning problems under the same imperative (i.e., minimisation of variational free energy) to solve a dual estimation problem. One subtlety here is that the free energy of the beliefs about parameters (i.e., the population density) minimises the path or time integral of free energy – during which the parameters are assumed not to change.

\subsection{Towards more flexible Expectations} Right now, our very direct way of hard-coding the expectations of our agent, in terms of explicit prior distributions on the first dimension of the agent's latent space, seems very ad-hoc and restricted. However, we could show that due to the flexibility of our agent's internal dynamics and the robust optimisation strategies, our agent quickly learns to reach these very narrowly defined goal and is able to build a realistic model of the environmental dynamics, which it encounters. In future work, we plan to look into more complex a priori beliefs. This could be done by sampling from the agent's generative model as part of the optimisation process. E.g. one could sample some processes from the generative model, calculate the quantity on which a constraint in terms of prior expectations of the agent should be placed, and calculate the difference between the sampled and the target distribution, e.g. using the KL-divergence. Then one could add this difference as penalty to the free energy. To enforce this constraint, one could use for example the Basic Differential Multiplier Method \citep{Platt1988}, which is similar to the use of Lagrange multipliers, but which can be used with gradient descent optimisation schemes. The idea is to add the penalty term $ P(\theta)$, which is equal to zero if the constraint is fulfilled, to the function to be minimised $ F(\theta)$, scaled by a multiplier $ \lambda$. The combined objective would look like this:

$$ F_\mathrm{constrained}(\theta) = F(\theta) + \lambda P(\theta)$$

To optimise the objective $ F(\theta)$  while forcing the penalty term $ P(\theta)$ to be zero, one can perform a gradient descent on $ F_\mathrm{constrained}(\theta)$ for the parameters $ \boldsymbol\theta$, but a gradient ascent for the penalty parameter $ \boldsymbol\lambda$. This prospective sampling to optimise the goals of the agent might be actually what parts of prefrontal cortex do when thinking about the future and how to reach certain goals.

\subsection{A first step from Artificial Life to Artificial General Intelligence} We present here a flexible and scalable implementation of a general active inference agent, that is able to learn a generative model of its environment while simultaneously achieving prespecified goals, in terms of prior expectations on its perceived states of the world. We hope that this implementation will prove useful to solve a wide variety of more complex and realistic problems. By this, one could show how general, intelligent behaviour follows naturally from the free energy principle. This principle, in turn, is derived from necessary properties of dynamic systems in exchange with changing and fluctuating environments, which allow them to sustain their identity by keeping their inner parameters within viable bounds, i.e. a very basic homeostatic argument \citep{Friston2012, Friston2013}. Thus, we hope that our work contributes to more concrete, experimental examples that intelligent behaviour necessary follows and is hard to separate from the basic imperative to survive in and adapt to changing environments. 

\section{Acknowledgements} The author would like to thank Karl Friston for his insightful comments on an earlier version of this manuscript and the participants and organisers of the Computational Psychiatry Course 2016 for stimulating lectures and discussions.

\bibliography{references}

\begin{thebibliography}{}

\bibitem[Adams et~al., 2013]{Adams2013}
Adams, R.~A., Stephan, K.~E., Brown, H., Frith, C.~D., and Friston, K.~J.
  (2013).
\newblock The computational anatomy of psychosis.
\newblock {\em Frontiers in Psychiatry}, 4(47).

\bibitem[Alais and Burr, 2004]{Alais2004}
Alais, D. and Burr, D. (2004).
\newblock The ventriloquist effect results from near-optimal bimodal
  integration.
\newblock {\em Current Biology}, 14(3):257--262.

\bibitem[Baez and Pollard, 2015]{Baez2015}
Baez, J.~C. and Pollard, B.~S. (2015).
\newblock Relative entropy in biological systems.
\newblock {\em https://arxiv.org/abs/1512.02742}.

\bibitem[Baltieri and Buckley, 2017]{Baltieri2017}
Baltieri, M. and Buckley, C.~L. (2017).
\newblock An active inference implementation of phototaxis.
\newblock {\em https://arxiv.org/abs/1707.01806}.

\bibitem[Berkes et~al., 2011]{Berkes2011}
Berkes, P., Orb{\'a}n, G., Lengyel, M., and Fiser, J. (2011).
\newblock Spontaneous cortical activity reveals hallmarks of an optimal
  internal model of the environment.
\newblock {\em Science}, 331:83--87.

\bibitem[Brockman et~al., 2016]{OpenAIGym}
Brockman, G., Cheung, V., Pettersson, L., Schneider, J., Schulman, J., Tang,
  J., and Zaremba, W. (2016).
\newblock Openai gym.
\newblock {\em https://arxiv.org/abs/1606.01540}.

\bibitem[Brown and Friston, 2012]{Brown2012}
Brown, H. and Friston, K.~J. (2012).
\newblock Free-energy and illusions: the cornsweet effect.
\newblock {\em Frontiers in Psychology}, 3(43).

\bibitem[Chung et~al., 2015]{Chung2015}
Chung, J., Kastner, K., Dinh, L., Goel, K., Courville, A., and Bengio, Y.
  (2015).
\newblock A recurrent latent variable model for sequential data.
\newblock {\em https://arxiv.org/abs/1506.02216}.

\bibitem[Conant and Ashby, 1970]{Ashby1970}
Conant, R. and Ashby, W. (1970).
\newblock Every good regulator of a system must be a model of that system.
\newblock {\em International Journal of Systems Science}, 1(2):89--97.

\bibitem[Ernst and Banks, 2002]{Ernst2002}
Ernst, M. and Banks, M. (2002).
\newblock Humans integrate visual and haptic information in a statistically
  optimal fashion.
\newblock {\em Nature}, 415:429--433.

\bibitem[Friston, 2005]{Friston2005}
Friston, K.~J. (2005).
\newblock A theory of cortical responses.
\newblock {\em Phil. Trans. R. Soc. B}, 360:815--836.

\bibitem[Friston, 2008]{Friston2008}
Friston, K.~J. (2008).
\newblock Hierarchical models in the brain.
\newblock {\em PLoS Computational Biology}, 4(11).

\bibitem[Friston, 2010]{Friston2010}
Friston, K.~J. (2010).
\newblock The free-energy principle: a unified brain theory?
\newblock {\em Nature Reviews Neuroscience}, 11(2):127--38.

\bibitem[Friston, 2012]{Friston2012}
Friston, K.~J. (2012).
\newblock A free energy principle for biological systems.
\newblock {\em Entropy}, 14:2100--2121.

\bibitem[Friston, 2013]{Friston2013}
Friston, K.~J. (2013).
\newblock Life as we know it.
\newblock {\em Journal of The Royal Society Interface}, 10.

\bibitem[Friston et~al., 2010]{FristonAl2010}
Friston, K.~J., Daunizeau, J., Kilner, J., and Kiebel, S.~J. (2010).
\newblock Action and behavior: a free-energy formulation.
\newblock {\em Biological Cybernetics}, 192(3):227--260.

\bibitem[Friston et~al., 2017]{Friston2017}
Friston, K.~J., Frith, C.~D., Pezzulo, G., Hobson, J.~A., and Ondobaka, S.
  (2017).
\newblock Active inference, curiosity and insight.
\newblock {\em Neural Computation}, 29:1--51.

\bibitem[Friston and Kiebel, 2009]{Friston2009}
Friston, K.~J. and Kiebel, S.~J. (2009).
\newblock Predictive coding under the free-energy principle.
\newblock {\em Phil. Trans. R. Soc. B}, 364:1211--1221.

\bibitem[Friston et~al., 2006]{FristonKilnerHarrison2006}
Friston, K.~J., Kilner, J., and Harrison, L. (2006).
\newblock A free energy principle for the brain.
\newblock {\em Journal of Physiology Paris}, 100:70--87.

\bibitem[Friston et~al., 2011]{Friston2011}
Friston, K.~J., Mattout, J., and Kilner, J. (2011).
\newblock Action understanding and active inference.
\newblock {\em Biological Cybernetics}, 104:137--160.

\bibitem[Friston et~al., 2015]{FristonEpistemicValue}
Friston, K.~J., Rigoli, F., Ognibene, D., Mathys, C., Fitzgerald, T., and
  Pezzulo, G. (2015).
\newblock Active inference and epistemic value.
\newblock {\em Cognitive Neuroscience}, 6(4):187--214.

\bibitem[Goodfellow et~al., 2016]{Goodfellow2016}
Goodfellow, I., Bengio, Y., and Courville, A. (2016).
\newblock {\em Deep Learning}.
\newblock MIT Press, \url{http://www.deeplearningbook.org}.

\bibitem[Haefner et~al., 2016]{Haefner2016}
Haefner, R., Berkes, P., and Fiser, J. (2016).
\newblock Perceptual decision-making as probabilistic inference by neural
  sampling.
\newblock {\em Neuron}, 90(3):649--660.

\bibitem[Harper, 2009]{Harper2009}
Harper, M. (2009).
\newblock The replicator equation as an inference dynamic.
\newblock {\em https://arxiv.org/abs/0911.1763}.

\bibitem[Hinton and Salakhutdinov, 2006]{Hinton2006}
Hinton, G.~E. and Salakhutdinov, R.~R. (2006).
\newblock Reducing the dimensionality of data with neural networks.
\newblock {\em Science}, 313.

\bibitem[Hornik et~al., 1989]{Hornik1989}
Hornik, K., Stinchcombe, M., and White, H. (1989).
\newblock Multilayer feedforward networks are universal approximators.
\newblock {\em Neural Networks}, 2:359--366.

\bibitem[Karpathy et~al., 2015]{Karpathy2015}
Karpathy, A., Johnson, J., and Fei-Fei, L. (2015).
\newblock Visualizing and understanding recurrent networks.
\newblock {\em https://arxiv.org/abs/1506.02078}.

\bibitem[Kingma and Ba, 2014]{BaKingma2014}
Kingma, D.~P. and Ba, J. (2014).
\newblock Adam: A method for stochastic optimization.
\newblock {\em https://arxiv.org/abs/1412.6980}.

\bibitem[Kingma et~al., 2016]{KingmaIAF}
Kingma, D.~P., Salimans, T., Jozefowicz, R., Chen, X., Sutskever, I., and
  Welling, M. (2016).
\newblock Improving variational inference with inverse autoregressive flow.
\newblock {\em https://arxiv.org/abs/1606.04934}.

\bibitem[Kingma and Welling, 2014]{KingmaWelling2014}
Kingma, D.~P. and Welling, M. (2014).
\newblock Auto-encoding variational bayes.
\newblock {\em ICLR}.

\bibitem[Knill and Pouget, 2004]{Knill2004}
Knill, D. and Pouget, A. (2004).
\newblock The bayesian brain: the role of uncertainty in neural coding and
  computation.
\newblock {\em Trends in Neurosciences}, 27(12):712--9.

\bibitem[Le et~al., 2015]{Hinton2015}
Le, Q.~V., Jaitly, N., and Hinton, G.~E. (2015).
\newblock A simple way to initialize recurrent networks of rectified linear
  units.
\newblock {\em https://arxiv.org/abs/1504.00941}.

\bibitem[LeCun et~al., 2015]{Lecun2015}
LeCun, Y., Bengio, Y., and Hinton, G.~E. (2015).
\newblock Deep learning.
\newblock {\em Nature}, 521:436--44.

\bibitem[Mnih et~al., 2015]{Mnih2015}
Mnih, V., Kavukcuoglu, K., Silver, D., Rusu, A.~A., Veness, J., Bellemare,
  M.~G., Graves, A., Riedmiller, M., Fidjeland, A.~K., Ostrovski, G., Petersen,
  S., Beattie, C., Sadik, A., Antonoglou, I., King, H., Kumaran, D., Wierstra,
  D., Legg, S., and Hassabis, D. (2015).
\newblock Human-level control through deep reinforcement learning.
\newblock {\em Nature}, 518:529--33.

\bibitem[Moore, 1991]{Moore1991}
Moore, A. (1991).
\newblock Variable resolution dynamic programming: Efficiently learning action
  maps in multivariate real-valued state-spaces.
\newblock In {\em Proceedings of the Eight International Conference on Machine
  Learning}. Morgan Kaufmann.

\bibitem[Moreno-Bote et~al., 2011]{Moreno-Bote2011}
Moreno-Bote, R., Knill, D., and Pouget, A. (2011).
\newblock Bayesian sampling in visual perception.
\newblock {\em Proc. Natl. Acad. Sci. U S A}, 108(30):12491--6.

\bibitem[Pathak et~al., 2017]{Pathak2017}
Pathak, D., Pulkit, A., Efros, A.~A., and Darrell, T. (2017).
\newblock Curiosity-driven exploration by self-supervised prediction.
\newblock {\em https://arxiv.org/abs/1705.05363}.

\bibitem[Platt and Barr, 1988]{Platt1988}
Platt, J.~C. and Barr, A.~H. (1988).
\newblock Constrained differential optimization.
\newblock In {\em Neural Information Processing Systems}, pages 612--621, New
  York. American Institute of Physics.

\bibitem[Rezende et~al., 2016]{Rezende2016}
Rezende, D.~J., Ali~Eslami, S.~M., Mohamed, S., Battaglia, P., Jaderberg, M.,
  and Heess, N. (2016).
\newblock Unsupervised learning of 3d structure from images.
\newblock {\em https://arxiv.org/abs/1607.00662}.

\bibitem[Rezende and Mohamed, 2015]{RezendeMohamed2015}
Rezende, D.~J. and Mohamed, S. (2015).
\newblock Variational inference with normalizing flows.
\newblock {\em JMRL}, 37.

\bibitem[Rezende et~al., 2014]{RezendeMohamedWierstra2014}
Rezende, D.~J., Mohamed, S., and Wierstra, D. (2014).
\newblock Stochastic backpropagation and approximate inference in deep
  generative models.
\newblock {\em ICML}.

\bibitem[Salimans et~al., 2017]{Salimans2017}
Salimans, T., Ho, J., Chen, X., and Sutskever, I. (2017).
\newblock Evolution strategies as a scalable alternative to reinforcement
  learning.
\newblock {\em https://arxiv.org/abs/1703.03864}.

\bibitem[Schwartenbeck et~al., 2015]{Schwartenbeck2015}
Schwartenbeck, P., Fitzgerald, T., Mathys, C., Dolan, R., Kronbichler, M., and
  Friston, K.~J. (2015).
\newblock Evidence for surprise minimization over value maximization in choice
  behavior.
\newblock {\em Scientific Reports}, 5(16575).

\bibitem[Siegelmann, 1995]{Siegelmann1995}
Siegelmann, H.~T. (1995).
\newblock Computation beyond the turing limit.
\newblock {\em Science}, 268:545--548.

\bibitem[{Theano Development Team}, 2016]{Theano2016}
{Theano Development Team} (2016).
\newblock {Theano: A {Python} framework for fast computation of mathematical
  expressions}.
\newblock {\em https://arxiv.org/abs/1605.02688}.

\bibitem[Todorov et~al., 2012]{Todorov2012}
Todorov, E., Erez, T., and Tassa, Y. (2012).
\newblock Mujoco: A physics engine for model-based control.
\newblock In {\em Proceedings of the IEEE/RSJ International Conference on
  Intelligent Robots and Systems (IROS)}.

\bibitem[Tomczak and Welling, 2016]{TomczakHHFlow}
Tomczak, J.~M. and Welling, M. (2016).
\newblock Improving variational auto-encoders using householder flow.
\newblock {\em https://arxiv.org/abs/1611.09630}.

\bibitem[Watson and Szathm{\'a}ry, 2016]{Watson2016}
Watson, R.~A. and Szathm{\'a}ry, E. (2016).
\newblock How can evolution learn?
\newblock {\em Trends in Ecology and Evolution}, 31(2):147--157.

\bibitem[Wong and Wang, 2006]{WongWang2006}
Wong, K.-F. and Wang, X.-J. (2006).
\newblock A recurrent network mechanism of time integration in perceptual
  decisions.
\newblock {\em The Journal of Neuroscience}, 26(4):1314--28.

\end{thebibliography}
\bibliographystyle{apalike}

\section{Supplementary Material}

\subsection{Performance without an explicit proprioceptive channel}

Even without direct feedback on its actions, in terms of a proprioceptive sensory channel $o_a$, our agent is able to successfully learn the goal instilled in terms of its prior expectations, while simultaneously building a generative model of its exteroceptive sensations. The convergence of the free energy bound is shown in supplementary~figure~\ref{fig_no_proprioception_convergence}. The true interaction of the agent with its environment after 30,000 training steps is shown in figure~\ref{fig_no_proprioception_propagated}, and its generative model of the world in figure~\ref{fig_no_proprioception_sampled}. The full code can be accessed at \url{http://www.github.com/kaiu85/deepAI_paper}.

\begin{figure}[h]
\makebox[\textwidth][c]{\includegraphics[width=1.0\textwidth]{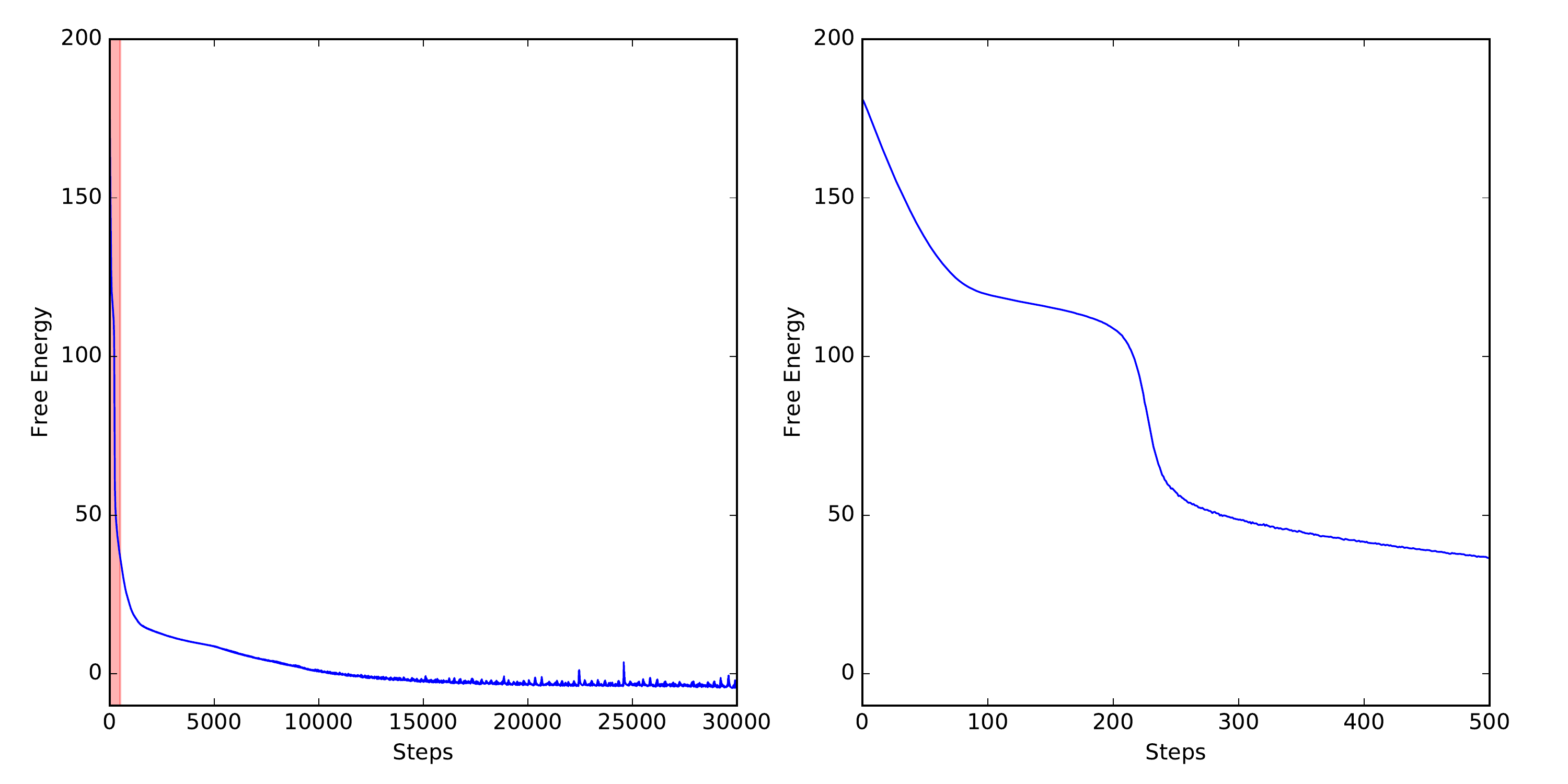}}
\caption{Convergence of an active inference agent, which does not possess a proprioceptive sensory channel, i.e. which gets no direct feedback on his actions. The area shaded in red in the left plot was enlarged in the right plot. }
\label{fig_no_proprioception_convergence}
\end{figure}

\begin{figure}[h]
\makebox[\textwidth][c]{\includegraphics[width=1.0\textwidth]{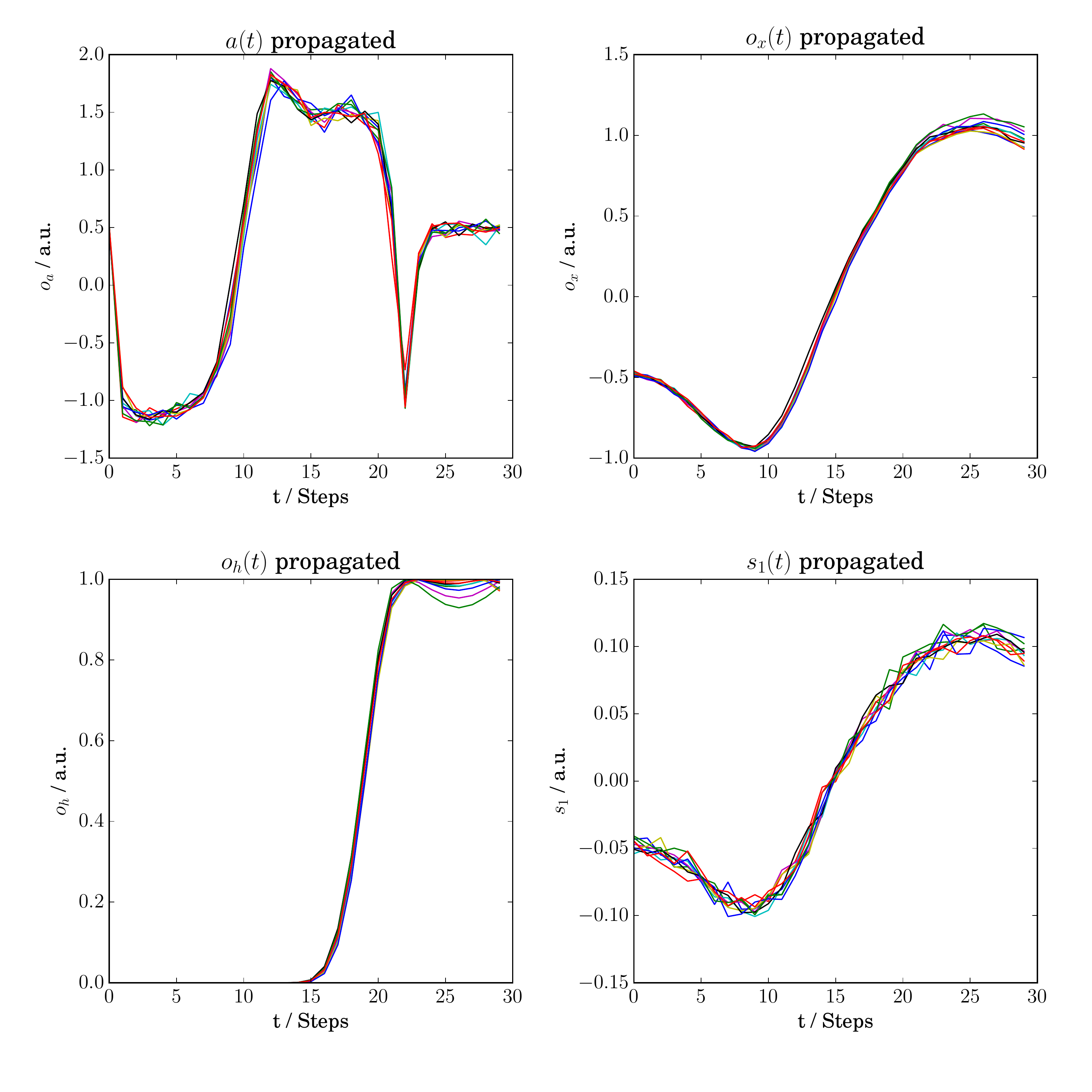}}
\caption{Performance of an agent without a proprioceptive sensory channel $o_a$ after 30,000 training steps, using the mean parameters of the population density. The agent has acquired a very efficient strategy to reach its goal position: it swings a bit to the left and then directly swings up to its goal position $x=1.0$. Shown are the agent's action $a$ (upper left), its sense of position $ o_x$ (upper right), its nonlinearly transformed sensory channel $ o_h$ and its "homeostatic" hidden state $ s_1$. }
\label{fig_no_proprioception_propagated}
\end{figure}

\begin{figure}[h]
\makebox[\textwidth][c]{\includegraphics[width=1.0\textwidth]{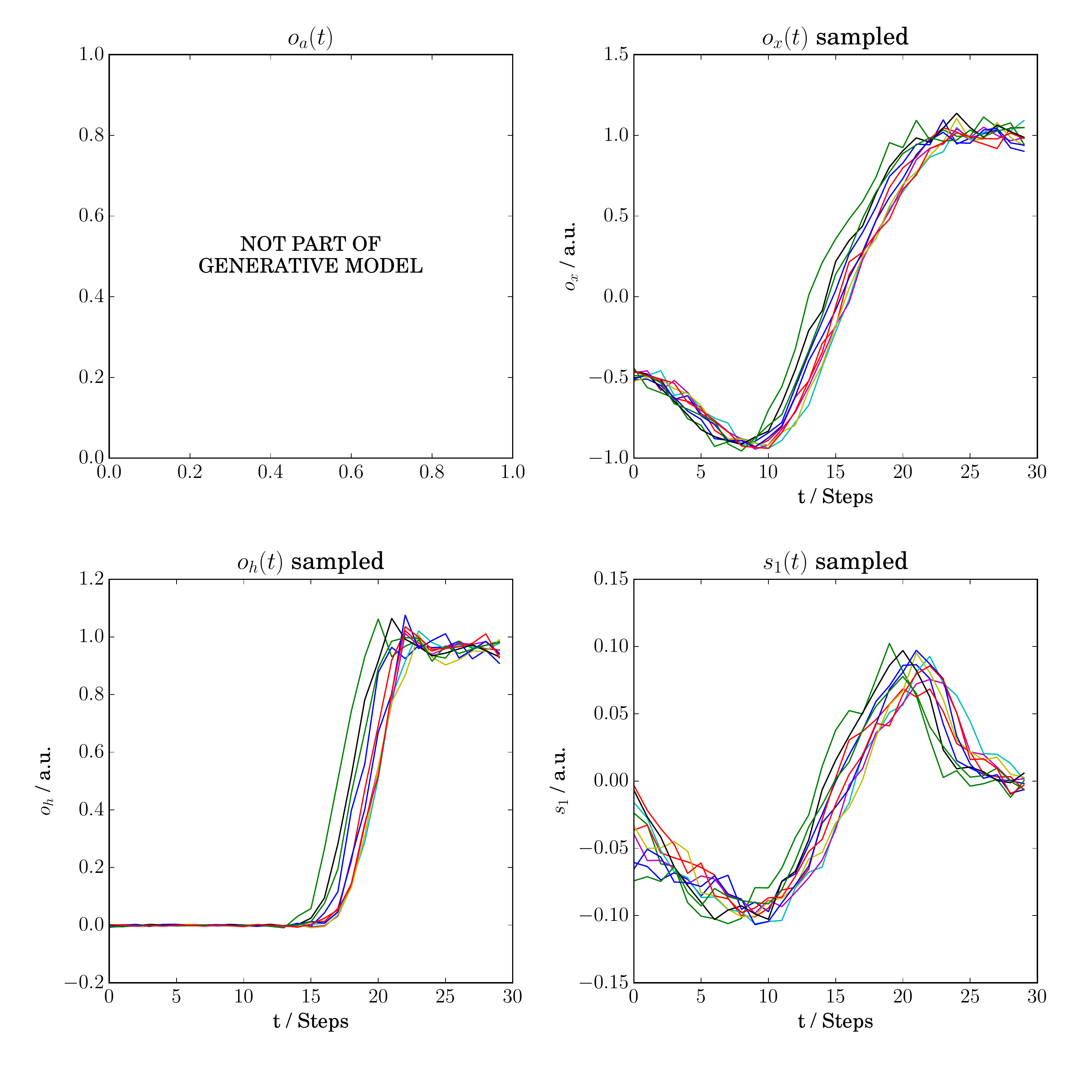}}
\caption{Sample from the generative model acquired by an agent which does not possess a proprioceptive sensory channel $o_a$ after 30,000 training steps. Shown are the agent's sense of position $ o_x$ (upper right), its nonlinearly transformed sensory channel $ o_h$ and its "homeostatic" hidden state $ s_1$. Note the very nice correspondence to its actual trajectory, when interacting with the world, as shown in supplementary figure~\ref{fig_no_proprioception_propagated} }
\label{fig_no_proprioception_sampled}
\end{figure}

\subsection{Performance with purely proprioceptive action}

If action is used only to directly suppress proprioceptive surprise, the convergence of the learning process is severely impaired, as shown in supplementary figure~\ref{fig_only_proprioceptive_action_convergence}. Here we optimise an agent whose structure, parameters and objective function are identical to the Deep Active Inference agent in the main text. However, the updates on the parameters of the agent's action function only depend on the gradient of the expectation value over the population density of the sensory surprise in the proprioceptive channel $<-\ln p_\theta(o_{a,t}|\mathbf{s}_t)>_{q(\mathbf{s}_t|\mathbf{s}_{t-1},o_{x,t},o_{h,t},o_{a,t})}$ with respect to the parameters of the action function. This corresponds to an agent which neglects the direct changes in other sensory modalities due to its actions.  One example might be the complex, nonlinear changes in the visual input to the retina, which arise even from small eye movements. The full code can be accessed at \url{http://www.github.com/kaiu85/deepAI_paper}.

Comparing supplementary figure~\ref{fig_only_proprioceptive_action_convergence} to figure~3 in the main text or to supplementary figure~1, which shows the convergence of an active inference agent lacking any proprioceptive input, it is obvious that this reduction prevents the agent from successfully achieving its goals and learning about its environment. This is also seen in the behavior of such an agent after 30,000 training steps, shown in supplementary figure~\ref{fig_only_proprioceptive_action_propagated}, and its - nonexistent - generative model of the world shown in supplementary figure~\ref{fig_only_proprioceptive_action_sampled}.

\begin{figure}
\makebox[\textwidth][c]{\includegraphics[width=1.0\textwidth]{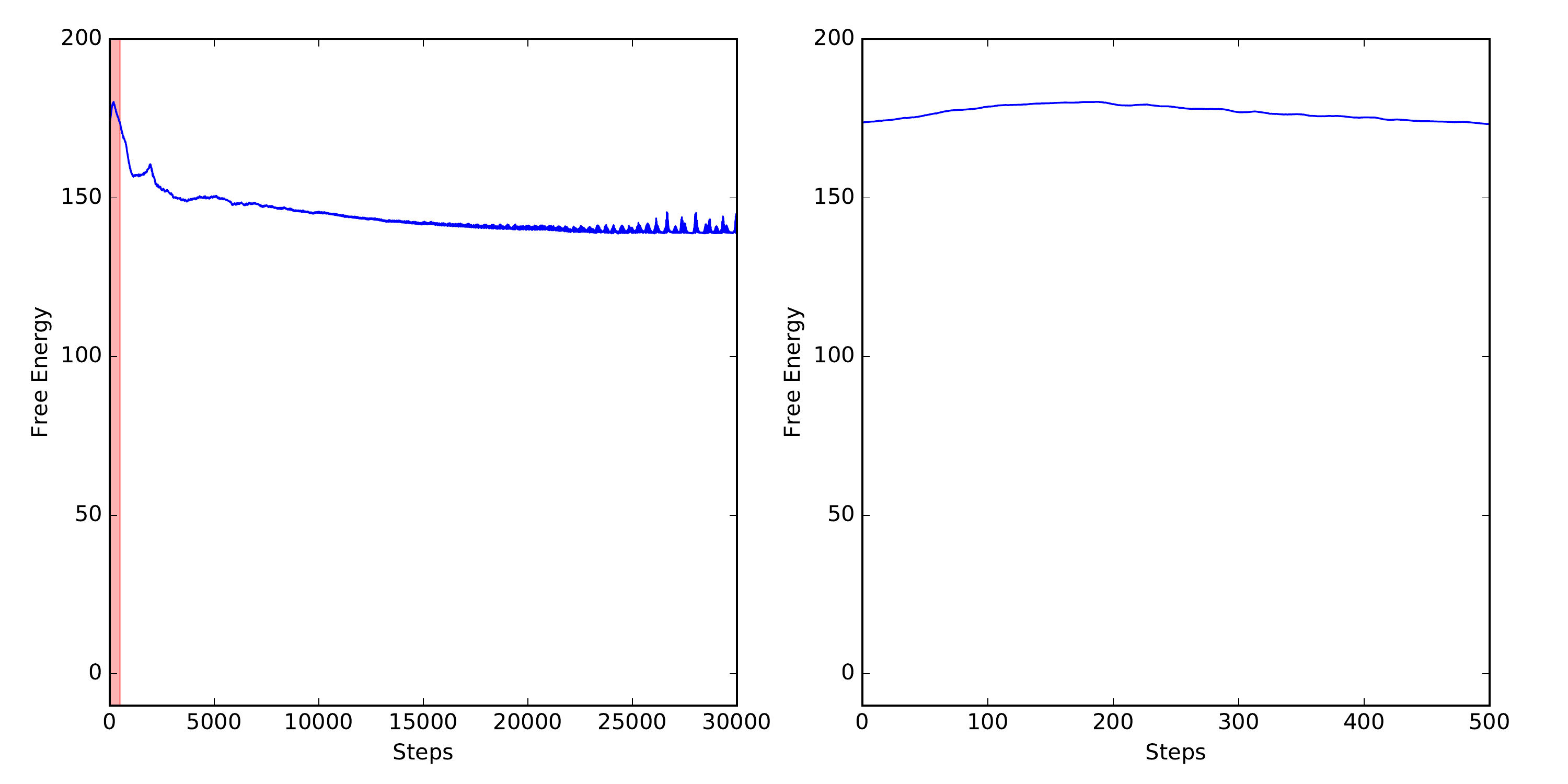}}
\caption{Convergence (or rather the lack of it) of an active inference agent, which uses action only to directly suppress its proprioceptive surprise. The area shaded in red in the left plot was enlarged in the right plot. }
\label{fig_only_proprioceptive_action_convergence}
\end{figure}

\begin{figure}
\makebox[\textwidth][c]{\includegraphics[width=1.0\textwidth]{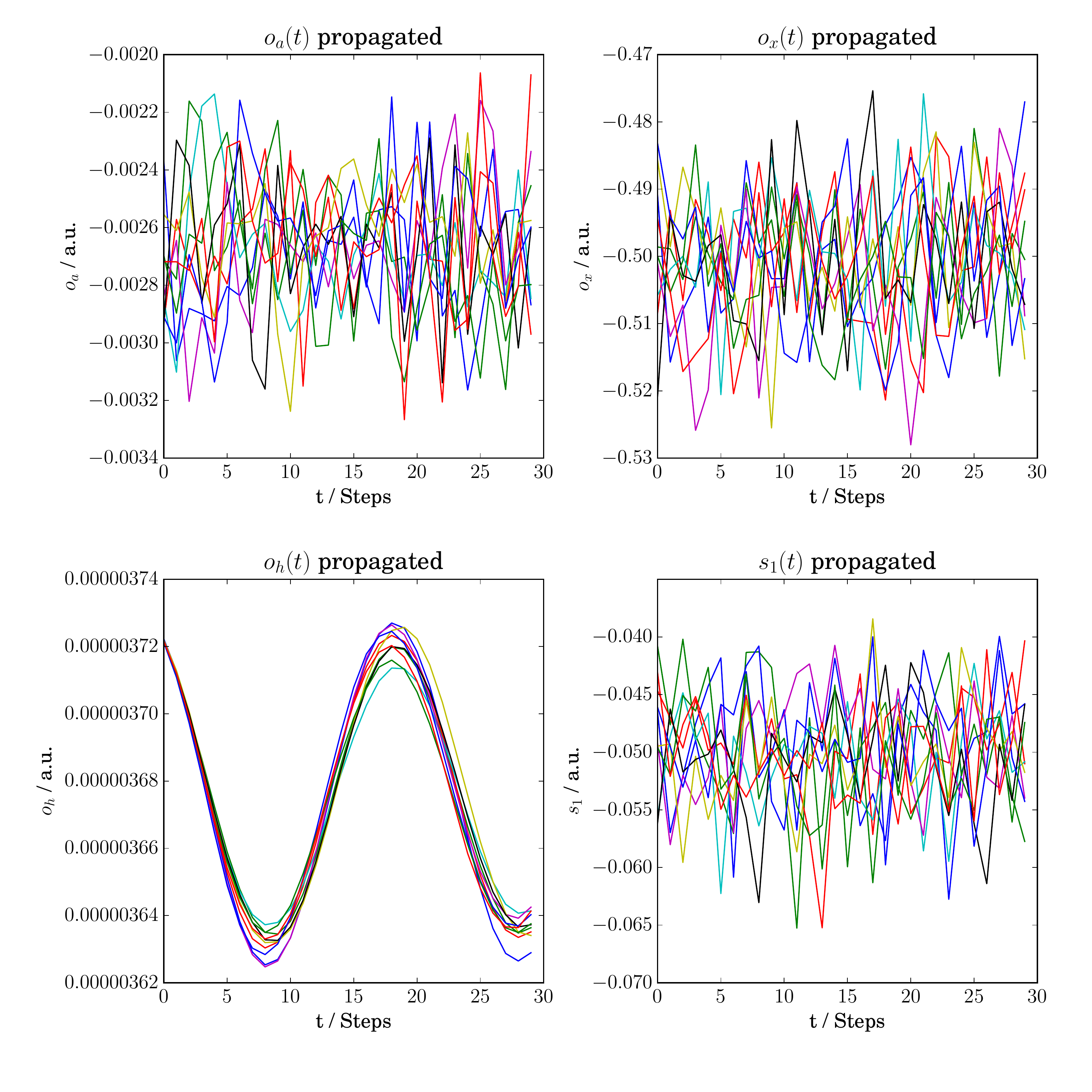}}
\caption{Performance of an agent which uses action only to directly suppress its proprioceptive surprise after 30,000 training steps, using the mean parameters of the population density. The agent is stuck at its initial position and shows no clear behavioral strategy. Shown are the agent's proprioceptive channel $o_a$ (upper left), its sense of position $ o_x$ (upper right), its nonlinearly transformed sensory channel $ o_h$ and its "homeostatic" hidden state $ s_1$. }
\label{fig_only_proprioceptive_action_propagated}
\end{figure}

\begin{figure}
\makebox[\textwidth][c]{\includegraphics[width=1.0\textwidth]{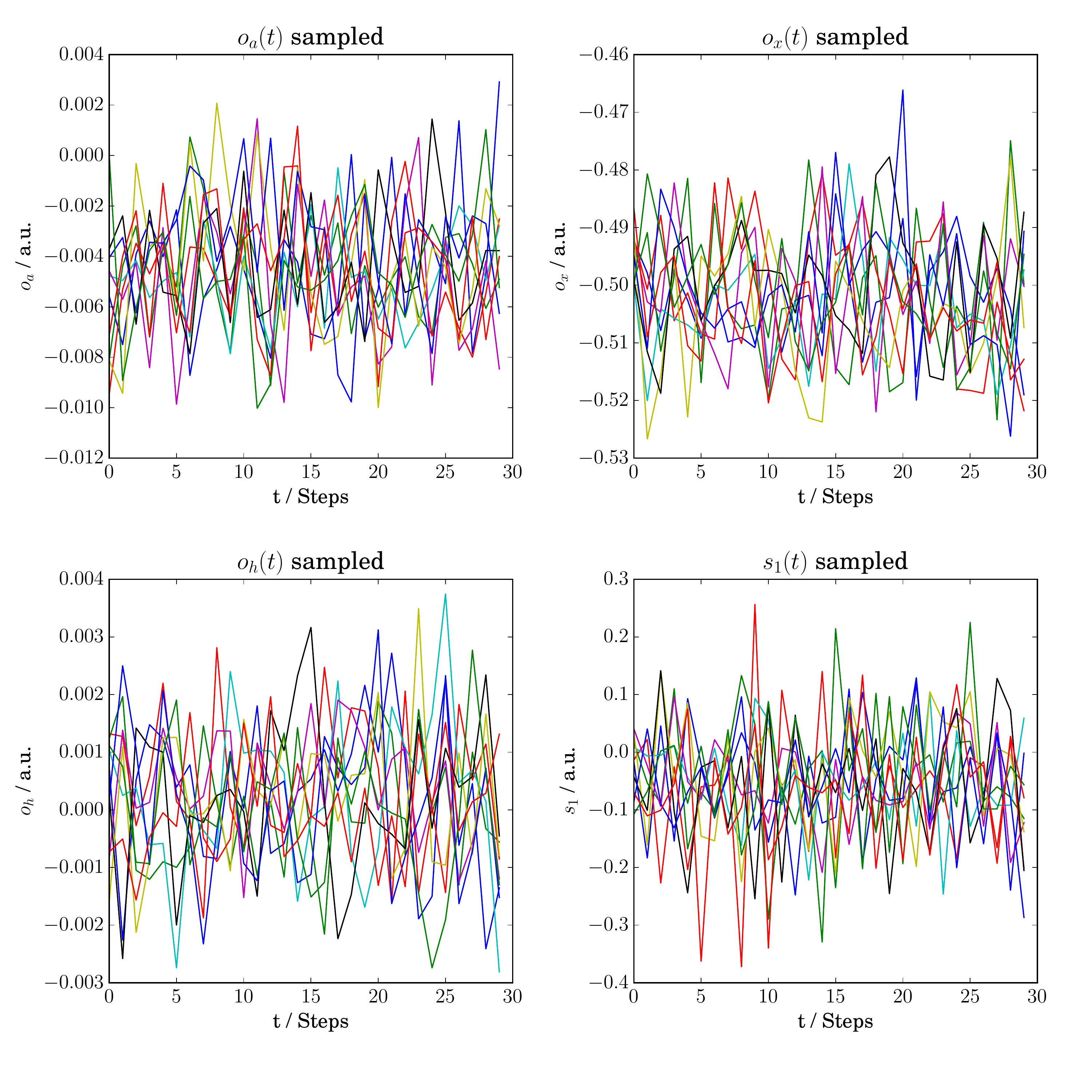}}
\caption{Sample from the generative model acquired by an agent which uses action only to directly suppress its proprioceptive surprise after 30,000 training steps. Shown are the agent's proprioceptive channel $o_a$ (upper left), its sense of position $ o_x$ (upper right), its nonlinearly transformed sensory channel $ o_h$ and its "homeostatic" hidden state $s_1$. Note the lacking correspondence to its actual trajectory (e.g. in $o_h$ or $s_1$), when interacting with the world, as shown in supplementary figure~\ref{fig_only_proprioceptive_action_propagated} }
\label{fig_only_proprioceptive_action_sampled}
\end{figure}

\end{document}